\documentclass[11pt,a4paper]{article}
\pdfoutput=1 %%need when submit to JHEP
\usepackage{jheppub}
\usepackage{tikz}
\newcommand{\ri}{\mathrm{i}}

\newcommand{\hf}{\frac{1}{2}}

\newcommand{\del}{\partial}

\newcommand{\h}[1]{\widehat{#1}}

\newcommand{\rt}[1]{\sqrt{#1}}
\newcommand{\bbC}{{\mathbb C}}
\newcommand{\bbZ}{{\mathbb Z}}
\newcommand{\bpartial}{{\bar\partial}}
\newcommand{\cS}{{\cal S}}
\newcommand{\hE}{\widehat{E}}
\newcommand{\tF}{\tilde{F}}
\newcommand{\cF}{{\cal F}}
\newcommand{\cZ}{{\cal Z}}
\newcommand{\btau}{{\bar{\tau}}}
\newcommand{\tD}{D_t}
\newcommand{\OPI}{{\mbox{\scriptsize 1PI}}}
\newcommand{\OPIF}{\cF^\OPI}
\newcommand{\OPR}{{\mbox{\scriptsize 1PR}}}
\newcommand{\OPRF}{\cF^\OPR}

\begin{document}

\title{Holomorphic anomaly of 2d Yang-Mills theory on a torus revisited}

\author[a]{Kazumi Okuyama}
\author[b]{and Kazuhiro Sakai}

\affiliation[a]{Department of Physics, 
Shinshu University, 3-1-1 Asahi, Matsumoto 390-8621, Japan}
\affiliation[b]{Institute of Physics, Meiji Gakuin University,
Yokohama 244-8539, Japan}

\emailAdd{kazumi@azusa.shinshu-u.ac.jp, kzhrsakai@gmail.com}

\abstract{We study the large $N$ 't Hooft expansion of the
chiral partition function of 2d $U(N)$ Yang-Mills theory on a torus.
There is a long-standing puzzle that no explicit
holomorphic anomaly equation is known for the partition function,
although it admits a topological string interpretation.
Based on the chiral boson interpretation
we clarify how holomorphic anomaly arises and propose
a natural anti-holomorphic deformation of the partition function.
Our deformed partition function obeys
a fairly traditional holomorphic anomaly equation.
Moreover, we find a closed analytic expression
for the deformed partition function.
We also study the behavior of the deformed partition function
both in the strong coupling/large area limit
and in the weak coupling/small area limit.
In particular, we observe that drastic simplification occurs
in the weak coupling/small area limit,
giving another nontrivial support for our anti-holomorphic deformation.
}

\maketitle

%%%%%%%%%%%%%%%%%%%%%%%%%%%%%%%%%%%%%%%%%%%%%%%%%%%%%%%%%%%%%%%%%%%%%%%%
\section{Introduction \label{sec:intro}}
%%%%%%%%%%%%%%%%%%%%%%%%%%%%%%%%%%%%%%%%%%%%%%%%%%%%%%%%%%%%%%%%%%%%%%%%

2d $U(N)$ Yang-Mills theory has been studied for a long time
as a prototypical example of exactly solvable gauge theories
\cite{Migdal:1975zg,Rusakov:1990rs,Witten:1992xu,Witten:1991we,
Blau:1991mp,Blau:1993hj}.
It has a profound connection with matrix models and
conformal field theories
\cite{Minahan:1993np,Douglas:1993wy,Dijkgraaf:1996iy}
as well as constitutes the integrable structures of
various field and string theory models, including topological strings
\cite{Cordes:1994sd,Cordes:1994fc,dijk,Li:2011mx},
supersymmetric gauge theories
\cite{Drukker:2007yx,Drukker:2007qr,Pestun:2009nn,Giombi:2009ek,
Gadde:2011ik,Tachikawa:2016kfc}
and black hole microstates
\cite{Vafa:2004qa,Aganagic:2004js,Dijkgraaf:2005bp}.

The large $N$ 't Hooft expansion of the
partition function of 2d $U(N)$ Yang-Mills theory
has been studied in the seminal paper by Gross and Taylor
\cite{Gross:1993hu}.
In the large $N$ limit the partition function
factorizes into the chiral and anti-chiral parts.
The chiral partition function $Z$ admits a string interpretation.
It has the characteristic form
$Z=\exp\bigl[\sum_{g=1}^\infty g_s^{2g-2}F_g\bigr]$
with the string coupling $g_s=1/N$,
where the free energy $F_g$ ``counts'' the maps of
a genus $g$ string world-sheet to the 2d target space.

In this paper we focus on the case where the target space
is a torus $T^2$.
The genus expansion of the partition function of
2d $U(N)$ Yang-Mills theory on a torus
has been studied in detail
\cite{Douglas:1993wy,Rudd:1994ta}.
In particular it was shown \cite{zagier,dijk} that
$F_g\ (g\ge 2)$ is a quasi modular form
for $SL(2,\bbZ)$ acting on the modulus $\tau$,
i.e.~$F_g$ is a polynomial of
the three Eisenstein series
$E_2(\tau),\,E_4(\tau),\,E_6(\tau)$.

It has been known that the chiral partition function
$Z$ is interpreted as the topological string partition function
for a class of certain non-compact Calabi-Yau threefolds
\cite{Vafa:2004qa}.
Concerning this there is a long-standing puzzle in the literature.
The puzzle is about holomorphic anomaly.
It is well known that
the topological string partition function
for Calabi-Yau threefolds
obeys the holomorphic anomaly equation \cite{Bershadsky:1993cx}.
In fact, it was proposed in \cite{Dijkgraaf:1996iy}
that $Z$ in the present case
satisfies the holomorphic anomaly equation of the form
\begin{align}
\label{eq:HAE-Dijkgraaf}
\frac{\partial Z}{\partial\bar\tau}
 =\left(\frac{g_s}{{\rm Im}\tau}\right)^2
  \frac{\partial^2 Z}{\partial\tau^2}.
\end{align}
In order to make sense of this equation,
one needs to restore the anti-holomorphic dependence on $\bar\tau$.
This is most commonly done by exploiting
the trade-off between the holomorphic anomaly
and the modular anomaly.
It is empirically known \cite{Minahan:1997ct,Grimm:2007tm} that 
in many cases
where the topological string amplitude is
expressed in terms of quasi modular forms in $\tau$,
the anti-holomorphic dependence on $\bar\tau$
is restored by merely replacing $E_2(\tau)$ with
\begin{align}
\hE_2(\tau,\bar\tau)=E_2(\tau)-\frac{3}{\pi{\rm Im}\tau}.
\label{eq:hat-E2}
\end{align}
However, it turns out that
\eqref{eq:HAE-Dijkgraaf} is actually not satisfied
when $\bar{\tau}$-dependence is restored in this way.
No correct holomorphic anomaly equation was found,
nor consistent recovery of $\bar{\tau}$-dependence was proposed
for the 2d Yang-Mills theory on a torus.\footnote{
For instance, there is a statement in 
\cite{Klemm:2015iya} that 
``In many cases no recursive holomorphic anomaly is known. E.g.~for the
2d QCD example it was argued in \cite{Rudd:1994ta} that such a recursion
does not exist.''
Also, in \cite{kanazawa}
it is stated that ``There is, however, no known explicit holomorphic
anomaly equations of higher genus for elliptic curves.''}

In this paper we resolve this puzzle by clarifying
how the holomorphic anomaly arises in the partition function.
Our construction is based on
the chiral boson interpretation of the 2d Yang-Mills theory on a torus.
It turns out that $E_2$'s are originated from
two kinds of source,
the propagator and the period integrals,
and only those from the former kind
is reasonably promoted to $\hE_2$.
Taking this into account,
we are able to identify the precise form of
the holomorphic anomaly equation.
We observe that the equation is of fairly traditional form
similar to \eqref{eq:HAE-Dijkgraaf},
but does not seem to be entirely equivalent.

We propose two different ways of
restoring the $\bar{\tau}$-dependence to $F=\ln Z$:
one is obtained as the free energy $\cF$ for all connected diagrams
and the other is obtained as the free energy $\OPIF$
for one-particle irreducible (1PI) diagrams only.
They are in fact related and
we find a relation which gives $\cF$ in terms of $\OPIF$.
Correspondingly, we obtain two holomorphic anomaly equations,
one is for $\cF$ and the other is for $\OPIF$.
The former is of traditional form as mentioned above while the latter
has a rather unconventional form.
Using the above relation between $\cF$ and $\OPIF$
we have verified
that the two holomorphic anomaly equations
are equivalent.

Moreover, we find a closed analytic expression
for the deformed partition function $\cZ=\exp\cF$.
Using this expression we show that 
the holomorphic anomaly equation is satisfied by $\cZ$
at all orders of the genus expansion.
We also study the behavior of the deformed partition function
when the modulus $t=-2\pi\ri\tau$ is large and small,
i.e.~the 't Hooft coupling and/or
the area of the torus is large and small.
In the limit of large $t$, $\cZ$ reduces to an Airy integral.
In the limit of small $t$, drastic simplification occurs
and $\cZ$ is given by a Fermi-Dirac integral.
This small $t$ behavior of $\cZ$ is even simpler than that of
the original partition function $Z$. We think that
this gives another nontrivial support
for our anti-holomorphic deformation.

This paper is organized as follows.
In section~\ref{sec:part},
we first review the free fermion representation
of the partition function without the anti-holomorphic dependence.
Next, we review the chiral boson interpretation and elucidate
how the anti-holomorphic dependence is naturally restored
to the partition function.
We then explain in detail how to calculate the free energy
with anti-holomorphic dependence using Feynman diagrams.
Finally we present the relation which expresses $\cF$
for connected diagrams in terms of $\OPIF$ for 1PI diagrams.
In section~\ref{sec:hae},
we present the holomorphic anomaly equations
for both $\cF$ and $\OPIF$.
We also give them a diagrammatic interpretation.
In section~\ref{sec:masterrep},
we present the closed analytic expression for
the deformed partition function $\cZ$.
We then study the large and small $t$ behavior of $\cZ$.
We also discuss how $\cF_g$ is determined
by using the holomorphic anomaly equation.
In section~\ref{sec:discussion},
we conclude with some discussion for future directions.
In Appendix~\ref{app:cF34},
we present a calculation of the free energy at $g=3,4$.
In Appendix~\ref{app:convention},
we summarize our convention of special functions
and present some useful relations.

%%%%%%%%%%%%%%%%%%%%%%%%%%%%%%%%%%%%%%%%%%%%%%%%%%%%%%%%%%%%%%%%%%%%%%%%
\section{\mathversion{bold}Partition function of 2d Yang-Mills theory
         on $T^2$\label{sec:part}}
%%%%%%%%%%%%%%%%%%%%%%%%%%%%%%%%%%%%%%%%%%%%%%%%%%%%%%%%%%%%%%%%%%%%%%%%

%%%
\subsection{Free fermion interpretation}
%%%

The chiral partition function of 2d $U(N)$ Yang-Mills theory on a torus
admits the large $N$ expansion\footnote{
Throughout this paper we ignore
the non-perturbative ${\cal O}(e^{-1/g_s})$ corrections.}
\begin{align}
Z(t)=\exp\left(\sum_{g=1}^\infty g_s^{2g-2}F_g(t)\right).
\label{eq:Z(t)}
\end{align}
Here, $g_s=1/N$ denotes the string coupling
and $t$ is the dimensionless combination
of the 't Hooft coupling and the area $A$ of the torus
\begin{align}
\label{eq:tdef}
t=g_{\rm YM}^2 N A.
\end{align}
The large $N$ 2d Yang-Mills theory
is often viewed as a string theory \cite{Gross:1992tu,Gross:1993hu}.
From this viewpoint $F_g(t)$
is regarded as the genus $g$ free energy,
which ``counts'' the maps of a genus $g$ string world-sheet
to the target space $T^2$. First few of them are
\begin{align}
\begin{aligned}
F_1&=-\ln\eta,\\
F_2&=\frac{5 E_2^3-3 E_2 E_4 -2 E_6}{51840},\\
F_3&=
 -\frac{6E_2^6-15E_2^4E_4-4E_2^3E_6+12E_2^2E_4^2+12E_2E_4E_6
        -7E_4^3-4E_6^2}
       {35831808},
\end{aligned}
\end{align}
where $\eta=\eta(\tau),\ E_{2n}=E_{2n}(\tau)$
are the Dedekind eta function and Eisenstein series of weight $2n$
respectively and
\begin{align}
%Q\equiv e^{2\pi i\tau}=e^{-t}.%,\qquad
\tau=\frac{\ri t}{2\pi}.
\end{align}

There are a number of ways to compute $F_g$.
Among others, it is worth mentioning that
the partition function $Z$ admits an interpretation in terms of
a system of non-relativistic free fermions on a circle
\cite{Minahan:1993np}.
This free fermion picture allows us to express $Z$
at large $N$ as \cite{Douglas:1993wy,dijk}
\begin{align}
\label{eq:Zfermirep}
Z&= Q^{-\frac{1}{24}}
 \oint\frac{dx}{2\pi\ri x}
 \prod_{p\in\bbZ_{\ge 0}+\hf}\left(1+xQ^p e^{g_s p^2/2}\right)
            \left(1+x^{-1}Q^p e^{-g_s p^2/2}\right),
\end{align}
where $Q= e^{2\pi\ri\tau}=e^{-t}$.
Based on this expression 
it was shown that $F_g\ (g\ge 2)$
is a quasi modular form of weight $6g-6$
for $SL(2,\bbZ)$ \cite{zagier,dijk}.
Using this fact one can, in principle,
compute $F_g$ up to any $g$ from the above expression.
Moreover, if we expand $Z$
as $Z=e^{F_1}(1+\sum_{n=1}^\infty g_s^{2n}Z_n)$,
$Z_n$ obeys a simple recursion relation,
originally found in \cite{zagier}
and improved in \cite{Okuyama:2018clk},
which determines $F_g$ much more efficiently
than just expanding \eqref{eq:Zfermirep}.

%%%
\subsection{Chiral boson interpretation}
%%%

As discussed in \cite{Douglas:1993wy,Dijkgraaf:1996iy}
(see also \cite{Cordes:1994fc,Li:2011mx}),
the bosonization maps
the free fermion system for the 2d Yang-Mills theory
to a theory of compactified boson field $\varphi$.
In this picture the chiral partition function is expressed as
\begin{align}
\label{eq:bosonicZ}
\cZ
=\int {\cal D}\varphi\exp\left[\int_{T^2}\left(
  \bpartial\varphi\partial\varphi
 +\frac{g_s}{3!}(\partial\varphi)^3\right)\right].
\end{align}
Here, we let a new symbol $\cZ$ denote
the ``bosonic'' (B-model) partition function,
meaning that it slightly differs from 
the ``fermionic'' (A-model) partition function $Z$
in \eqref{eq:Zfermirep}.
As is well known, the correspondence between
the fermionic and bosonic pictures
is nontrivial and
the exact equivalence of the partition functions
is achieved only if one appropriately takes account of all the
winding modes of the compactified boson $\varphi$.
In defining $\cZ$, however,
we consider only the local quantum fluctuation
and ignore the winding mode contribution.
$\cZ$ then deviates from $Z$ and
gets mild $\btau$ dependence.
We will see that
$\cZ$ defined in this way provides us with a natural generalization
of $Z$.

We expand the partition function \eqref{eq:bosonicZ} as
\begin{align}
\ln\cZ
 =\cF
 =\sum_{g=1}^\infty g_s^{2g-2}\cF_g.
\label{eq:sum-g}
\end{align}
The free energy $\cF_g$ is then given by
\begin{align}
\label{eq:Fgasexpvalue}
\cF_g
 =\left\langle \left(\frac{1}{3!}\int(\partial\varphi)^3\right)^{2g-2}
  \right\rangle_{\rm connected}.
\end{align}
The expectation value is evaluated by the Wick contraction
by means of the propagator
\begin{align}
G(z_1,z_2):=\langle\partial\varphi(z_1)\partial\varphi(z_2)\rangle.
\end{align}
We take the torus here (i.e.~on the B-model side) to be
\begin{align}
\label{eq:B-modelT2}
T^2:=\ \bbC/(2\pi\bbZ+2\pi\tau\bbZ).
\end{align}
The propagator is then given by
\cite{Douglas:1993wy,Dijkgraaf:1996iy}
\begin{align}
\begin{aligned}
G(z_1,z_2)
 &=-\wp(z_1-z_2)-\frac{\hE_2}{12}\\
&=-\wp(z_1-z_2)-\frac{E_2}{12}+S.
\end{aligned}
\label{eq:prop_z}
\end{align}
Here, $\hE_2$ is defined in \eqref{eq:hat-E2}
and we have introduced the notation
\begin{align}
\label{eq:Sdef}
S:=\frac{1}{t+\bar{t}}=\frac{1}{4\pi{\rm Im}\tau}=\frac{E_2-\hE_2}{12}.
\end{align}
The propagator \eqref{eq:prop_z} is
obtained by taking two derivatives
of the usual free boson propagator
$\langle\varphi(z_1,\bar z_1)\varphi(z_2,\bar z_2)\rangle$.
Note that $S$ accounts for the background charge, which is
inversely proportional to
the area $4\pi^2{\rm Im}\tau$ of the torus \eqref{eq:B-modelT2}.
Such a non-holomorphic piece may or may not remain
in the final result, depending on
one's treatment of the winding mode of $\varphi$.
If one takes account of all the winding mode contributions,
$S$ is eventually canceled out \cite{Datta:2014zpa}.
On the other hand, if one restricts oneself to
the large $N$ limit,
the compact nature of $\varphi$ is not seen \cite{Douglas:1993wy}
and one can consistently ignore the winding mode.
We choose the latter option and keep the $S$ dependence intact.

The free energy \eqref{eq:Fgasexpvalue} is computed as the sum of
all possible connected Feynman diagrams with $2g-2$ trivalent
vertices.
To evaluate the Feynman diagrams, we need to specify
the integration prescription.
We follow the prescription of
\cite{Douglas:1993wy, Boehm:2013}:
For a Feynman diagram $\Gamma$
with $2g-2$ trivalent vertices,
the integral is given by
\begin{align}
\label{eq:IGamma}
I_\Gamma
 =\oint\prod_{i=1}^{2g-2}\frac{dx_i}{2\pi\ri x_i}
  \prod_{k=1}^{3g-3}G(x_k^+/x_k^-)
 ={\rm Coeff}_{x_1^0\cdots x_{2g-2}^0}
  \left[\prod_{k=1}^{3g-3}G(x_k^+/x_k^-)\right],
\end{align}
where $x_i$ labels the $i$th vertex and $x_k^\pm$ are the two vertices
connected by the $k$th propagator.
The propagator $G(x)$ is given by
\begin{align}
\label{eq:prop_x}
\begin{aligned}
G(x)&=\sum_{n=1}^\infty\frac{n(x^n+x^{-n}Q^n)}{1-Q^n}+S\\
    &=\sum_{n\in\bbZ}\frac{xQ^n}{(1-xQ^n)^2}+S,
 \end{aligned}
\end{align}
which is equivalent to $G(z,0)$ given in \eqref{eq:prop_z}
with the identification $x=e^{\ri z}$.\footnote{
By abusing notation
we let the same propagator be denoted by $G(z,0)$ or $G(x)$,
depending on the context.}
As demonstrated in \cite{Boehm:2013},
one can easily evaluate the Feynman diagram at least
in the small $Q$ expansion.
One difference from \cite{Boehm:2013} is that
here we have the non-holomorphic piece $S$ in the propagator.
Consequently, the result is not simply a quasi modular form,
but a polynomial in $S$ whose coefficients are
given by quasi modular forms.
More specifically, we assume that the result takes the form
\begin{align}
\label{eq:IGamma-ansatz}
I_\Gamma = \sum_{k=0}^{3g-3} S^k f_{6g-6-2k}(\tau)
\end{align}
with $f_{2n}(\tau)$ being a quasi modular form of weight $2n$.
Then we can find the exact expression
by matching the small $Q$ expansion.

A few technical remarks are in order:
The residue integral in \eqref{eq:IGamma}
is equivalent to
the period integral
over one of the fundamental cycles of the torus.
In terms of the variable $z$, the period integral is expressed as
\begin{align}
\oint dz(\cdots) =\frac{1}{2\pi}\int_{z_0}^{z_0+2\pi}dz(\cdots),
\label{eq:A-peri}
\end{align}
where $z_0$ is a generic value
chosen in such a way that the integration path avoids the singularities.
For instance, it is easy to check that
the above prescription based on \eqref{eq:IGamma} and \eqref{eq:prop_x}
reproduces
the period integrals used in \cite{Douglas:1993wy}
\begin{align}
\label{eq:pintwp}
\oint dz\wp(z)=-\frac{E_2}{12},\qquad
\oint dz\wp(z)^2=\frac{E_4}{144},\qquad
\oint dz\wp(z)^3=\frac{4E_6-9E_2E_4}{8640}.
\end{align}
Next, the evaluation of the multiple Laurent series
expansion \eqref{eq:IGamma} 
needs care, because
the order of the expansions matters.
The reader is referred for the details to
the reference \cite{Boehm:2013}.
Here we simply mention that
what we need to do is to take the symmetric average of all
the possible orderings of $|x_i|$ for the position $x_i$ of vertices
in \eqref{eq:IGamma}.
Third, the symmetry factor
is simply given by the order of the automorphism
group of the graph $\Gamma$ \cite{Boehm:2013}.
This means that
the contribution of the Feynman diagram $\Gamma$ to the free energy is
to be normalized as
\begin{align}
\label{eq:symfactor}
\frac{1}{|{\rm Aut}(\Gamma)|}I_\Gamma.
\end{align}

Due to the presence of $S$ in the propagator
a new feature appears:
1-particle reducible (1PR) diagrams do not vanish
and yield nontrivial contribution to the free energy.
One can argue that the integral $I$ for the
1PR diagram $\Gamma$ always has the structure
$I=SI_1I_2$, as follows:
Suppose that $\Gamma$ can be decomposed as $\Gamma_1\cup\Gamma_2$
with a single propagator $G(x_a/x_b)$
connecting $\Gamma_1$ and $\Gamma_2$.
The contribution of $\Gamma$ is written schematically as
\begin{align}
\begin{aligned}
I&=\oint\frac{dx_a}{2\pi\ri x_a}\frac{dx_b}{2\pi\ri x_b}G(x_a/x_b)\\
 &\hspace{1em}\times
  \prod_{u\in\Gamma_1}\frac{dx_u}{2\pi\ri x_u}
  \prod G(x_{u_i}/x_{u_j})G(x_{u_k}/x_a)
  \prod_{v\in\Gamma_2}\frac{dx_v}{2\pi\ri x_v}
  \prod G(x_{v_i}/x_{v_j})G(x_{v_k}/x_b).
\end{aligned}
\label{eq:1PRint1}
\end{align}
Since the propagator depends only on the ratio of $x$'s,
we can set $x_a=x_b=1$ in the second line of \eqref{eq:1PRint1}
by rescaling $x_u\to x_u x_a$ and $x_v\to x_v x_b$.
Thus we find
\begin{align}
\label{eq:Ifactorization}
\begin{aligned}
I&=\oint\frac{dx_a}{2\pi\ri x_a}\frac{dx_b}{2\pi\ri x_b} G(x_a/x_b)\\
 &\hspace{1em}\times
  \prod_{u\in\Gamma_1}\frac{dx_u}{2\pi\ri x_u}
    \prod G(x_{u_i}/x_{u_j})G(x_{u_k})
  \prod_{v\in\Gamma_2}\frac{dx_v}{2\pi\ri x_v}
    \prod G(x_{v_i}/x_{v_j})G(x_{v_k})\\
 &=SI_1I_2,
\end{aligned}
\end{align}
where $I_k\ (k=1,2)$ is the integral for the diagram $\Gamma_k$.
From this
it is clear that all 1PR diagrams vanish
in the limit $\btau\to-\ri\infty$ or $S\to 0$.
This is why one has only to consider
1PI diagrams in the literature.

1PR diagrams often contain the self-contraction
$G(x/x)=G(z,z)$.
Naively, it diverges as
$-\wp(\epsilon)\sim -1/\epsilon^2$ for $\epsilon\to 0$.
However, the self-contraction always appears
as the tadpole diagram and thus
the amplitude can be consistently renormalized.
We do this by regularizing $G(z,z)$ as
\begin{align}
\label{eq:selfcont}
G(z,z)=-\frac{E_2}{12}+S=-\frac{\hE_2}{12}.
\end{align}
One can understand it by applying
the zeta-function regularization
to \eqref{eq:prop_x}:\footnote{See \cite{Li:2011mi}
for another regularization using the heat kernel.}
The propagator $G(x)$ in \eqref{eq:prop_x} at $x=1$ is regularized as
\begin{equation}
\begin{aligned}
 G(x=1)&=\sum_{n=1}^\infty n+2\sum_{n=1}^\infty\frac{nQ^n}{1-Q^n}+S\\
&=\zeta(-1)+\frac{1-E_2}{12}+S=-\frac{E_2}{12}+S.
\end{aligned} 
\end{equation}
Here we have used $\zeta(-1)=-1/12$.

An important identity, which we will use frequently, is
\begin{align}
\label{eq:DGrel}
DG(z_1,z_2)=\oint dz_3G(z_1,z_3)G(z_3,z_2),
\end{align}
where
\begin{align}
\label{eq:Ddef}
D :=-\partial_t=Q\partial_Q
  =\frac{1}{2\pi\ri}\frac{\partial}{\partial\tau}.
\end{align}
This identity can easily be shown
by using the expression \eqref{eq:prop_x} as
\begin{align}
\begin{aligned}
&\hspace{-2em}
\oint\frac{dy}{2\pi\ri y}G(y/x_1)G(y/x_2)\\
 &=S^2+\oint\frac{dy}{2\pi\ri y}
   \sum_{n,m=1}^\infty
   \frac{n(y^n/x_1^n+(x_1^n/y^n)Q^n)}{1-Q^n}
   \frac{m(y^m/x_2^m+(x_2^m/y^m)Q^m)}{1-Q^m}\\
 &=S^2
   +\sum_{n=1}^\infty\frac{n^2(x_1^n/x_2^n+x_2^n/x_1^n)Q^n}{(1-Q^n)^2}\\
 &=DG(x_1/x_2).
\end{aligned}
\end{align}
Here we have used $DS=S^2$ in \eqref{eq:diffformulas}.
Note that the identity \eqref{eq:DGrel}
has a simple diagrammatic representation,
as shown in Figure~\ref{Fig:DGrel}.
\begin{figure}[t]
\vspace{4ex}
\begin{equation*}
D\ 
\begin{tikzpicture}
\draw[line width=1pt](0,0)--(1.6,0);
\node at (0,.35){$1$};
\node at (1.6,.35){$2$};
\filldraw(0,0)circle(.08);
\filldraw(1.6,0)circle(.08);
\end{tikzpicture}
\quad =\quad 
\begin{tikzpicture}
\draw[line width=1pt](0,0)--(1.6,0);
\node at (0,.35){$1$};
\node at (.8,.35){$3$};
\node at (1.6,.35){$2$};
\filldraw(0,0)circle(.08);
\filldraw(.8,0)circle(.08);
\filldraw(1.6,0)circle(.08);
\end{tikzpicture}
\end{equation*}
\vspace{-4ex}
\caption{Diagrammatic representation of the identity \eqref{eq:DGrel}.
\label{Fig:DGrel}}
\end{figure}
That is, the differential operator $D$
inserts an internal vertex into a propagator.
Note also that the regularization \eqref{eq:selfcont}
is compatible with the identity \eqref{eq:DGrel}.

\begin{figure}[t]
\vspace{4ex}
\begin{equation*}
\begin{array}{c@{\hspace{3em}}c@{\hspace{3em}}c}
\begin{tikzpicture}
\draw[line width=1pt](.5,.5)circle(.5);
\end{tikzpicture}
&
\begin{tikzpicture}
\draw[line width=1pt](.5,.5)circle(.5);
\filldraw(0,.5)circle(.08);
\end{tikzpicture}
&
\begin{tikzpicture}
\draw[line width=1pt](.5,.5)circle(.5);
\filldraw(0,.5)circle(.08);
\filldraw(1,.5)circle(.08);
\end{tikzpicture}
\\
\cF_1&D\cF_1&D^2\cF_1
\end{array}
\end{equation*}
\caption{Feynman diagrams for $\cF_1$ and its derivatives.
\label{Fig:FeynmanF1}}
\end{figure}
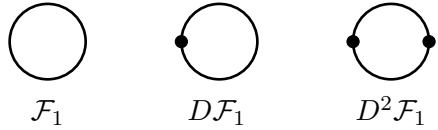

The form of $\cF_1$ is not computed from the Feynman diagram,
but it is rather related to the normalization of the path integral
\eqref{eq:bosonicZ}.
It is convenient to fix it as
\begin{align}
\label{eq:cF1}
\cF_1&=\ln\sqrt{2\pi S}-\ln\eta.
\end{align}
Its derivatives $D\cF_1$ and $D^2\cF_1$ are computed
easily  as
\begin{align}
\label{eq:diffcF1}
D\cF_1=\frac{1}{2}S-\frac{E_2}{24},\qquad
D^2\cF_1=\frac{1}{2}S^2-\frac{E_2^2-E_4}{288}.
\end{align}
(See \eqref{eq:diffformulas} for the differentiation
of $S,\,\eta$ and $E_{2n}$.)
Fixed in this way, $\cF_1$ and its derivatives
have a nice diagrammatic interpretation,
as shown in Figure~\ref{Fig:FeynmanF1}:
One can formally regard a circle as the Feynman diagram for $\cF_1$.
As the differential operator $D$ inserts
an internal vertex into a propagator,
the Feynman diagram for
$D\cF_1$ and $D^2\cF_1$ should then be a circle with
one and two internal vertex(ices) respectively.
Indeed, one can express \eqref{eq:diffcF1} also as
\begin{align}
\label{eq:diffcF1Grep}
D\cF_1
 =\frac{1}{2}\oint dz G(z,z),\qquad
D^2\cF_1
 =\frac{1}{2}\oint dz_1dz_2 G(z_1,z_2)^2,
\end{align}
which are, up to normalization,
the very integrals derived from the Feynman diagrams.\footnote{
One might think that
the symmetry factor for $D^2\cF_1$ would be $1/4$
according to the rule \eqref{eq:symfactor},
but it is actually $1/2$ as we see in \eqref{eq:diffcF1Grep}.
This is not contradictory since the rule \eqref{eq:symfactor}
is derived for diagrams made up of trivalent vertices only
and does not necessarily apply to the present case.}
The expressions \eqref{eq:diffcF1Grep} are verified easily by using
\eqref{eq:selfcont} and \eqref{eq:pintwp},
or the latter expression is derived from the former one
by using \eqref{eq:DGrel}.

Using the above techniques,
one can compute $\cF_g$ up to any $g$ in principle.
As an illustration let us compute $\cF_2$.
\begin{figure}[t]
\vspace{4ex}
\begin{equation*}
\begin{array}{c@{\hspace{2.5em}}c}
\begin{tikzpicture}
\draw[line width=1pt](0,.5)--(1,.5);
\draw[line width=1pt](.5,.5)circle(.5);
\filldraw(0,.5)circle(.08);
\filldraw(1,.5)circle(.08);
\end{tikzpicture}
&
\raisebox{.2cm}{
\begin{tikzpicture}
\draw[line width=1pt](.6,.5)--(1.2,0.5);
\draw[line width=1pt](.3,.5)circle(.3);
\draw[line width=1pt](1.5,.5)circle(.3);
\filldraw(.6,.5)circle(.08);
\filldraw(1.2,.5)circle(.08);
\end{tikzpicture}
}
\\[.5ex]
\cF_2^{(1)}&\cF_2^{(2)}
\end{array}
\end{equation*}
\caption{Feynman diagrams for $\cF_2$.\label{Fig:Feynman2}}
\end{figure}
As shown in Figure~\ref{Fig:Feynman2},
there are two Feynman diagrams that contribute to $\cF_2$.
We therefore express $\cF_2$ as
\begin{align}
\cF_2=\cF_2^{(1)}+\cF_2^{(2)}
\end{align}
with
\begin{align}
\begin{aligned}
\cF_2^{(1)}
 &:=\frac{1}{12}\oint dz_1 dz_2 G(z_1,z_2)^3,\\
\cF_2^{(2)}
 &:=\frac{1}{8}\oint dz_1 dz_2 G(z_1,z_1)G(z_1,z_2)G(z_2,z_2).
\end{aligned}
\end{align}
First, $\cF_2^{(1)}$ is evaluated as
\begin{align}
\label{eq:F21eval}
\begin{aligned}
\cF_2^{(1)}
 &=\frac{1}{12}\oint dz_1dz_2 G(z_1,0)^3
  =\frac{1}{12}\oint dz_1 G(z_1,0)^3\\
 &=-\frac{1}{12}\oint dz_1
 \left[\frac{\hE_2^3}{12^3}
 +3\frac{\hE_2^2}{12^2}\wp(z_1)
 +3\frac{\hE_2}{12}\wp(z_1)^2
 +\wp(z_1)^3
 \right]\\
 &=\frac{-5\hE_2^3+15E_2\hE_2^2-15E_4\hE_2+9E_2E_4-4E_6}{103680}\\
 &=\frac{1}{12}S^3-\frac{E_2^2-E_4}{576}S
     +\frac{5E_2^3-3E_2E_4-2E_6}{51840}.
\end{aligned}
\end{align}
In the first equality we have changed
the integration variables as $(z_1,z_2)\to (z_1+z_2,z_2)$
and in the third line we have used \eqref{eq:pintwp}.
The final form has been appeared previously
in the literature \cite{Datta:2014zpa}.
Note that 
$\cF_2^{(1)}$ becomes $F_2$ in the limit $S\to 0$,
reproducing the original calculation of \cite{Douglas:1993wy}.
Next, let us evaluate $\cF_2^{(2)}$.
This is easy because the diagram for
$\cF_2^{(2)}$ is 1PR and thus
the integral is factorized
as in \eqref{eq:Ifactorization}.
It is clear from the Feynman diagram that
\begin{align}
\label{eq:F22eval}
\begin{aligned}
\cF_2^{(2)}
 &=\frac{1}{2}S(D\cF_1)^2
  =\frac{1}{2}S\left(\frac{1}{2}S-\frac{E_2}{24}\right)^2.
\end{aligned}
\end{align}
We thus obtain
\begin{align}
\cF_2
 &=\frac{5}{24}S^3-\frac{E_2}{48}S^2
  -\frac{E_2^2-2E_4}{1152}S+\frac{5E_2^3-3E_2E_4-2E_6}{51840}.
\label{eq:cF2-result}
\end{align}
We also present the results of $\cF_3$ and $\cF_4$
in Appendix~\ref{app:cF34}.

As one can see from the above calculation, 
there are two sources of $E_2$ or $\hE_2$:
one is the explicit dependence of $\hE_2$
in the propagator \eqref{eq:prop_z} and the other $E_2$
is coming from the 
integration of $\wp(z)$ in \eqref{eq:pintwp}. 
The latter does not appear in the combination of $\hE_2=E_2-12S$ and
thus the final result of $\cF_g$ is a mixture of $E_2$ and $\hE_2$ (see 
e.g.~the third line of \eqref{eq:F21eval}).
This clearly shows that the naive prescription of replacing all $E_2$ by
$\hE_2$ does not work in the case of 2d Yang-Mills theory.

One can choose any linear combination of $S$ and $E_2$ as a basis
when writing $\cF_g$ as a polynomial.
In \eqref{eq:cF2-result} we expressed $\cF_2$ in terms of the basis
$\{S,E_2,E_4,E_6\}$.
Another convenient basis is $\{S,\hE_2,E_4,E_6\}$,
in which $\cF_2$ is written as
\begin{equation}
\begin{aligned}
\cF_2=\frac{5\hE_2^2+2E_4}{1920}S+\frac{5\hE_2^3-3\hE_2E_4-2E_6}{51840}.
\end{aligned} 
\label{eq:cF2-hE2}
\end{equation}
From this one can see that $\cF_g$ does not come back to itself under
the modular transformation since $S$ transforms inhomogeneously
(see \eqref{eq:modular}
for the modular transformation of $\{S,\hE_2,E_4,E_6\}$).
This lack of modularity can be understood from the definition of
$\cF_g$: as shown in \eqref{eq:IGamma} and \eqref{eq:A-peri},
the Feynman diagrams are evaluated as period integrals along
the A-cycle $z\in[0,2\pi]$ of $T^2$,
and the A-cycle and the B-cycle $z\in[0,2\pi\tau]$ are treated 
asymmetrically in our formalism.

%%%
\subsection{1PI free energy}
%%%

The free energy $\cF_g\ (g\ge 2)$ is calculated by evaluating
all connected diagrams.
One can decompose $\cF_g$ as
\begin{align}
\cF_g= \OPIF_g+\OPRF_g,
\end{align}
where $\OPIF_g$ and $\OPRF_g$ are the contributions
from the 1PI and 1PR diagrams respectively.
As we saw above,
all 1PR diagrams vanish when $S=0$,
namely $\OPRF_g(t,S=0)=0$.
On the other hand, we have defined $\cF_g$ so that
$\cF_g(t,S=0)=F_g(t)$. Therefore, we have
\begin{align}
\OPIF_g(t,S=0)=F_g(t)\quad(g\ge 2).
\end{align}
This implies that not only $\cF_g$ but also
$\OPIF_g$ can be regarded as a natural
anti-holomorphic deformation
of $F_g$.

The explicit form of $\OPIF_g$ can be easily obtained for small $g$.
The first two are
\begin{align}
\begin{aligned}
\OPIF_1&=\cF_1\hspace{.7em}
 =\ln\sqrt{2\pi S}-\ln\eta,\\
\OPIF_2&=\cF_2^{(1)}
 =\frac{1}{12}S^3-\frac{E_2^2-E_4}{576}S
 +\frac{5E_2^3-3E_2E_4-2E_6}{51840}.
\end{aligned}
\end{align}
We also present the results of
$\OPIF_3$ and $\OPIF_4$ in Appendix~\ref{app:cF34}.

The factorization property \eqref{eq:Ifactorization}
of 1PR Feynman integrals
suggests that $\cF_g$ is written in terms of
$\OPIF_{g'}$ with $g'\le g$.
Indeed, one observes that
\begin{align}
\begin{aligned}
\cF_1&=\OPIF_1,\\
\cF_2&=\OPIF_2+\frac{1}{2}S(D\OPIF_1)^2,\\
\cF_3&=\OPIF_3+SD\OPIF_1 D\OPIF_2+\frac{1}{2}S^2D^2\OPIF_1(D\OPIF_1)^2
 +\frac{1}{6}S^3(D\OPIF_1)^3.
\end{aligned}
\end{align}
To describe the general rule,
let us introduce the total 1PI and 1PR free energies
\begin{align}
\OPIF:=\sum_{g=1}^\infty g_s^{2g-2}\OPIF_g,\qquad
\OPRF:=\sum_{g=1}^\infty g_s^{2g-2}\OPRF_g.
\end{align}
We conjecture that $\cF$ is expressed in terms of $\OPIF$ as
\begin{align}
\label{eq:F-F1PIrel}
\cF(t,S)=\lim_{\hbar\to 0}\hbar\ln
 \left[\frac{1}{\sqrt{2\pi\hbar S}}
   \int_{-\infty}^\infty d\phi
   \exp\left(\frac{1}{\hbar}\left[
      -\frac{\phi^2}{2S}+\frac{g_s\phi^3}{6}
      +\OPIF(t-g_s\phi,S)\right]\right)
 \right].
\end{align}
This expression
should be understood by means of the power series expansion
in $g_s$. Each coefficient of the expansion
is evaluated essentially by the Gaussian integral.

The relation \eqref{eq:F-F1PIrel}
has a simple interpretation.
To see this, we subtract $\OPIF$ from both
sides of the equation and rewrite it as
\begin{align}
\label{eq:F1PR-F1PIrel}
\begin{aligned}
&\hspace{-.5em}
\OPRF(t,S)\\
&=\lim_{\hbar\to 0}\hbar\ln
 \left[\frac{1}{\sqrt{2\pi\hbar S}}
   \int_{-\infty}^\infty d\phi
   \exp\left(\frac{1}{\hbar}\left[
      -\frac{\phi^2}{2S}+\frac{g_s\phi^3}{6}
      +\sum_{n=1}^\infty\frac{g_s^n\phi^n}{n!}D^n\OPIF(t,S)
\right]\right)
 \right].
\end{aligned}
\end{align}
Let us now regard the ordinary integral in $\phi$
as a ``path integral.''
We then see that the ``action'' consists of
the quadratic kinetic term,
the cubic interaction and the $n$-point interactions.
The free energy given by the above ``path integral''
is calculated by
evaluating all connected diagrams made up of these interaction
vertices connected by the propagator $S$.
The limit $\hbar\to 0$ means that
we have only to consider tree-level diagrams.
Indeed, in the original boson theory
any 1PR diagram is a tree graph
whose vertices are the trivalent vertex
or the $n$-point 1PI diagrams with $n>0$.
The edge of the tree graph is given by the propagator $G$,
which in this case reduces to $S$ due to
the factorization property \eqref{eq:Ifactorization}.
Note that the symmetry factor for a graph is
simply given by the inverse of the order of its automorphism group,
in the same way as in \eqref{eq:symfactor}.

As we will see, the relation \eqref{eq:F-F1PIrel}
is very stimulating in regard to understanding
about our main result presented in section~\ref{sec:masterrep}.

%%%%%%%%%%%%%%%%%%%%%%%%%%%%%%%%%%%%%%%%%%%%%%%%%%%%%%%%%%%%%%%%%%%%%%%%
\section{Holomorphic anomaly equation\label{sec:hae}}
%%%%%%%%%%%%%%%%%%%%%%%%%%%%%%%%%%%%%%%%%%%%%%%%%%%%%%%%%%%%%%%%%%%%%%%%

A long-standing puzzle in the literature is that
the partition function $Z$ does not seem to satisfy
a simple holomorphic anomaly equation.
As we mentioned,
the puzzle originates in the wrong assumption that
anti-holomorphic derivative $\partial_{\btau}$
is essentially equivalent to $\partial_{E_2}$.
This is empirically true for many cases,
but does not apply to the present case.
As we saw in the last section, there are two sources of $E_2$'s
and only $E_2$'s brought by the propagator
should be replaced with $\hE_2$.
In this section we will see that the free energy $\cF_g$
and the partition function $\cZ$
indeed satisfy a usual holomorphic anomaly equation.
The form of the equation is slightly different from
the original proposal \eqref{eq:HAE-Dijkgraaf}
in \cite{Dijkgraaf:1996iy}.
We will also present a holomorphic anomaly equation for $\OPIF_g$.

%%%
\subsection{Holomorphic anomaly equation for connected free energy}
%%%

By using the explicit form of $\cF_g$,
it is not difficult to write down the holomorphic anomaly equation
for small $g$.
Since $\btau$ always appears through $S$,
we can use
\begin{align}
\label{eq:diffSdifftau}
\partial_S = 8\pi\ri({\rm Im}\tau)^2\partial_\btau
\end{align}
as the anti-holomorphic derivative.

It follows immediately from
\eqref{eq:diffcF1}, \eqref{eq:F21eval} and \eqref{eq:F22eval}
that
\begin{align}
\partial_S\cF_2^{(1)}=\frac{1}{2}D^2\cF_1,\qquad
\partial_S\cF_2^{(2)}=\frac{1}{2}(D\cF_1)^2+\frac{1}{2}SD\cF_1.
\end{align}
From this one can write the holomorphic anomaly equation
for $\cF_2$ as
\begin{align}
2\partial_S\cF_2=(D+S)D\cF_1+D\cF_1D\cF_1.
\end{align}
One can repeat the same calculation for $\cF_3$ and $\cF_4$.
The results are
\begin{align}
\begin{aligned}
2\partial_S\cF_3&=(D+S)D\cF_2+2D\cF_1D\cF_2,\\
2\partial_S\cF_4&=(D+S)D\cF_3+2D\cF_1D\cF_3+D\cF_2D\cF_2.
\end{aligned}
\end{align}
Regarding these results we conjecture that
\begin{align}
\label{eq:HAE_Fg}
2\partial_S\cF_g=(D+S)D\cF_{g-1}+\sum_{h=1}^{g-1}D\cF_hD\cF_{g-h}
\quad (g\ge 2).
\end{align}
We will present several consistency checks of this equation.

Let us first mention that
the equation \eqref{eq:HAE_Fg}
admits a natural diagrammatic interpretation
as shown in Figure~\ref{Fig:HAE}.
Recall that $S$ always comes into the free energy
through the propagator \eqref{eq:prop_z}.
Therefore, the action of $\partial_S$
is interpreted as
the removal of a propagator from the Feynman diagram.
Diagrammatically, there are three different cases:
$D^2\cF_{g-1}$ and $SD\cF_{g-1}$
on the right hand side of \eqref{eq:HAE_Fg}
correspond to removal of a normal propagator and
a self-contracted one respectively,
where the resulting diagram remains connected.
$D\cF_hD\cF_{g-h}$ corresponds to removal of a normal propagator,
where the resulting diagram becomes disconnected.
\begin{figure}[t]
\vspace{4ex}
\begin{equation*}
\begin{array}{ccccccc}
\raisebox{.525cm}{$2\partial_S\ $}
\begin{tikzpicture}
\draw[fill=lightgray,line width=1pt](.6,.6)circle(.6);
\node at (.6,.6){$\cF_g$};
\end{tikzpicture}
&
\raisebox{.525cm}{$\quad =\quad $}
&
\begin{tikzpicture}
\draw[dotted,line width=1pt](.3,.6)circle(.3);
\draw[fill=lightgray,line width=1pt](.9,.6)circle(.6);
\node at (.9,.6){$\cF_{g-1}$};
\filldraw(.375,.3095)circle(.08);
\filldraw(.375,.8905)circle(.08);
\end{tikzpicture}
&
\raisebox{.525cm}{$\quad +\quad $}
&
\begin{tikzpicture}
\draw[line width=1pt](.6,.6)--(1.2,0.6);
\draw[dotted,line width=1pt](.3,.6)circle(.3);
\draw[fill=lightgray,line width=1pt](1.8,.6)circle(.6);
\node at (1.8,.6){$\cF_{g-1}$};
\filldraw(.6,.6)circle(.08);
\filldraw(1.2,.6)circle(.08);
\end{tikzpicture}
&
\raisebox{.525cm}{$\quad+\quad\displaystyle\sum_{h=1}^{g-1}\ $}
&
\begin{tikzpicture}
\draw[dotted, line width=1pt](1.2,.6)--(1.8,0.6);
\draw[fill=lightgray,line width=1pt](.6,.6)circle(.6);
\draw[fill=lightgray,line width=1pt](2.4,.6)circle(.6);
\node at (.6,.6){$\cF_h$};
\node at (2.4,.6){$\cF_{g-h}$};
\filldraw(1.2,.6)circle(.08);
\filldraw(1.8,.6)circle(.08);
\end{tikzpicture}
\\[.5em]
&&D^2\cF_{g-1}&&SD\cF_{g-1}&&D\cF_hD\cF_{g-h}
\end{array}
\end{equation*}
\caption{Diagrammatic interpretation of
the holomorphic anomaly equation \eqref{eq:HAE_Fg}.
The propagator removed by the action of $\partial_S$
is indicated by the dotted line.
\label{Fig:HAE}}
\end{figure}

It is also possible to express 
the holomorphic anomaly equation \eqref{eq:HAE_Fg}
in terms of $\cF$ or $\cZ$ in \eqref{eq:sum-g}.
One can easily show that \eqref{eq:HAE_Fg}  is equivalent to
\begin{align}
\label{eq:HAE_F}
2\partial_S\cF -S^{-1}
 = g_s^2\left[(D+S)D\cF +(D\cF)^2\right],
\end{align}
where we have subtracted the genus-one term $2\del_S\cF_1=S^{-1}$
on the left hand side.
One can also recast \eqref{eq:HAE_F} into the equation
for the partition function $\cZ=\exp \cF$
\begin{align}
\label{eq:HAE_Z}
\left(2\partial_S-S^{-1}\right)\cZ = g_s^2(D+S)D\cZ.
\end{align}
We can remove the awkward term $S^{-1}$ on the left hand side of
\eqref{eq:HAE_Z}
by rescaling the partition function as
\begin{equation}
\begin{aligned}
 \h{\cZ}=\frac{\eta}{\rt{2\pi S}}\cZ
 =\exp\left(\sum_{g=2}^\infty g_s^{2g-2}\cF_g \right).
\end{aligned} 
\label{eq:hatZ}
\end{equation}
Then the holomorphic anomaly equation for $\h{\cZ}$ becomes
\begin{equation}
\begin{aligned}
 2\del_S\h{\cZ}=g_s^2\left(D+S-\frac{\hE_2}{24}\right)
\left(D-\frac{\hE_2}{24}\right)\h{\cZ}. 
\end{aligned} 
\end{equation}
One can rewrite it in terms of $\partial_\btau$
using \eqref{eq:diffSdifftau}.
Written in this form, our holomorphic anomaly equation
is similar to the original proposal \eqref{eq:HAE-Dijkgraaf}
of \cite{Dijkgraaf:1996iy},
but does not seem to be entirely equivalent.

Our holomorphic anomaly equation is very reminiscent of
that of Bershadsky-Cecotti-Ooguri-Vafa (BCOV)
for Calabi-Yau threefolds \cite{Bershadsky:1993cx}.
It was shown that topological string amplitude $\tF_g\ (g\ge 2)$
for any Calabi-Yau threefold
is a polynomial in the generators
$\cS^{ij}$, $\cS^i$, $\cS$, $K_i$ \cite{Yamaguchi:2004bt,Alim:2007qj}.
By regarding $\tF_g$ as a function in these generators
$\tF_g(\cS^{ij},\cS^i,\cS,K_i;z_i,\bar{z}_i)$,
the BCOV holomorphic anomaly equation is written as
\cite{Alim:2007qj, Alim:2015qma}
\begin{align}
\label{AlimLange1}
\frac{\partial\tF_g}{\partial\cS^{ij}}
 &=\frac{1}{2}\sum_{h=1}^{g-1}D_i\tF_hD_j\tF_{g-h}
  +\frac{1}{2}D_iD_j\tF_{g-1},\\
\label{AlimLange2}
0&=\frac{\partial\tF_g}{\partial K_i}
  +\cS^i\frac{\partial\tF_g}{\partial\cS}
  +\cS^{ij}\frac{\partial\tF_g}{\partial\cS^j}.
\end{align}
By identifying the coordinate and the propagator as
\begin{align}
z^1=t,\qquad
\cS^{11}=S
\end{align}
and the covariant derivatives as
\begin{align}
D_1\tF_g=D\cF_g,\qquad
D_1D_1\tF_{g-1}=(D+S)D\cF_{g-1},
\end{align}
\eqref{AlimLange1} coincides with our equation \eqref{eq:HAE_Fg}.
However, this is merely a heuristic argument and it should not be taken
at face value since we have not computed the connection and the
covariant derivative on the moduli space.
The detail of this calculation can be found in \cite{Hosono:2008np}.
It would be interesting to understand the more precise relation between
our holomorphic anomaly equation \eqref{eq:HAE_Fg}
and that of BCOV \eqref{AlimLange1}.

%%%
\subsection{Holomorphic anomaly equation for 1PI free energy}
%%%

In this subsection we see that
$\OPIF_g$ also obeys a simple holomorphic anomaly equation.
For small $g$ one can explicitly derive that
\begin{align}
\begin{aligned}
2\partial_S\OPIF_1&=S^{-1},\\
2\partial_S\OPIF_2&=D^2\OPIF_1,\\
2\partial_S\OPIF_3&=D^2\OPIF_2+S\left(D^2\OPIF_1\right)^2,\\
2\partial_S\OPIF_4
 &=D^2\OPIF_3+2SD^2\OPIF_2D^2\OPIF_1+S^2\left(D^2\OPIF_1\right)^3.
\end{aligned}
\end{align}
We conjecture that these relations follow
from the holomorphic anomaly equation of the form
\begin{align}
\label{eq:HAE_1PIF}
2S\partial_S\OPIF = \frac{1}{1-g_s^2SD^2\OPIF}.
\end{align}
By means of power series expansion in $g_s$
we have checked that \eqref{eq:HAE_1PIF}
is equivalent to \eqref{eq:HAE_F}
under the identification \eqref{eq:F-F1PIrel}.
It would be interesting to prove this equivalence.

The equation \eqref{eq:HAE_1PIF}
also has a simple diagrammatic interpretation.
To see this, we rewrite it as
\begin{align}
\label{eq:HAE_1PIFbis}
2\partial_S\OPIF
 =S^{-1}+g_s^2D^2\OPIF+g_s^4S\left(D^2\OPIF\right)^2
  +g_s^6S^2\left(D^2\OPIF\right)^3+\cdots.
\end{align}
Written in this form, the equation
is interpreted as follows:
The first term on the right hand side
accounts for $2\partial_S\OPIF_1=S^{-1}$ at order ${\cal O}(g_s^0)$.
At order ${\cal O}\bigl(g_s^{2g-2}\bigr)\ (g\ge 2)$,
$\partial_S$ acting on a 1PI diagram removes a propagator from it.
The resulting diagram is a connected diagram
and thus viewed as a tree graph whose vertices
are 1PI diagrams or the trivalent vertex.
In order for the original diagram to be 1PI, however,
the tree graph in this case cannot have any branching.
That is, the resulting diagram is always a linear graph on
1PI components as shown in Figure~\ref{Fig:HAE1PI}.
This represents \eqref{eq:HAE_1PIFbis}.
\begin{figure}[t]
\vspace{4ex}
\begin{equation*}
\begin{array}{cccccc}
\raisebox{.525cm}{$2\partial_S\ $}
\begin{tikzpicture}
\draw[fill=lightgray,line width=1pt](.6,.6)circle(.6);
\node at (.6,.65){$\OPIF$};
\end{tikzpicture}
&
\raisebox{.525cm}{$\quad =\quad S^{-1}\quad+$}
&
\begin{tikzpicture}
\tikzset{
    partial ellipse/.style args={#1:#2:#3}{
        insert path={+ (#1:#3) arc (#1:#2:#3)}
    }
}
\draw[dotted,line width=1pt](.9,.98)
     [partial ellipse=-60.4:240.4:1.2 and .45];
\draw[fill=lightgray,line width=1pt](.9,.6)circle(.6);
\node at (.9,.65){$\OPIF$};
\filldraw(.3,.6)circle(.08);
\filldraw(1.5,.6)circle(.08);
\end{tikzpicture}
&
\raisebox{.525cm}{$+\hspace{-.3em}$}
&
\begin{tikzpicture}
\tikzset{
    partial ellipse/.style args={#1:#2:#3}{
        insert path={+ (#1:#3) arc (#1:#2:#3)}
    }
}
\draw[dotted,line width=1pt] (1.5,1.05)
     [partial ellipse=-61:241:2.4 and .6];
\draw[fill=lightgray,line width=1pt](.6,.6)circle(.6);
\draw[fill=lightgray,line width=1pt](2.4,.6)circle(.6);
\node at (.6,.65){$\OPIF$};
\node at (2.4,.65){$\OPIF$};
\filldraw(0,.6)circle(.08);
\filldraw(1.2,.6)circle(.08);
\filldraw(1.8,.6)circle(.08);
\filldraw(3,.6)circle(.08);
\draw[line width=1pt](1.2,.6)--(1.8,.6);
\end{tikzpicture}
&
\raisebox{.525cm}{$\hspace{-.3em}+\quad\cdots$}
\\[.5em]
&&g^2_sD^2\OPIF&&g^4_sS\left(D^2\OPIF\right)^2&
\end{array}
\end{equation*}
\caption{Diagrammatic interpretation of
the holomorphic anomaly equation \eqref{eq:HAE_1PIFbis}.
The propagator removed by the action of $\partial_S$
is indicated by the dotted line.
\label{Fig:HAE1PI}}
\end{figure}

We note that \eqref{eq:HAE_1PIF} can also be written as
\begin{align}
\label{eq:HAE_1PIFg}
\partial_S\OPIF_g= \sum_{h=1}^{g-1}S\partial_S\OPIF_h D^2\OPIF_{g-h}
\qquad(g\ge 2).
\end{align}
This is obtained by multiplying both sides of \eqref{eq:HAE_1PIF}
by the factor $\left(1-g_s^2SD^2\OPIF\right)$
and then expanding in $g_s$.
This expression is convenient for practical use.

%%%%%%%%%%%%%%%%%%%%%%%%%%%%%%%%%%%%%%%%%%%%%%%%%%%%%%%%%%%%%%%%%%%%%%%%
\section{Master representation and general properties
         \label{sec:masterrep}}
%%%%%%%%%%%%%%%%%%%%%%%%%%%%%%%%%%%%%%%%%%%%%%%%%%%%%%%%%%%%%%%%%%%%%%%%

The holomorphic anomaly equation allows us to compute $\cF_g$ 
efficiently up to very high order.
This enables us to find an all-order expression for $\cZ$,
as presented below.
Using this expression we discuss some general properties of $\cZ$.

%%%
\subsection{Master representation}
%%%

We conjecture that the partition function $\cZ(t,S)$
defined by the boson theory is simply given by
\begin{align}
\label{eq:masterformula}
\cZ(t,S)
 =\int_{-\infty}^\infty d\phi\,
 e^{-\frac{\phi^2}{2S}+\frac{g_s\phi^3}{6}}Z(t-g_s\phi).
\end{align}
Here, $Z(t)$ is the partition function defined in the fermion theory
\eqref{eq:Zfermirep}.
The expression \eqref{eq:masterformula}
should be understood by means of the power series expansion
in $g_s$: each coefficient of the expansion
is evaluated essentially by the Gaussian integral.

Interestingly,
the above relation between $\cZ$ and $Z$
is very similar to \eqref{eq:F-F1PIrel}
that relates $\cF$ and $\OPIF$.
Regarding this similarity and rewriting \eqref{eq:masterformula} as
\begin{align}
\label{eq:masterformula2}
e^{\cF(t,S)}
 =\int_{-\infty}^\infty d\phi\,
 \exp\left(
 -\frac{\phi^2}{2S}+\frac{g_s\phi^3}{6}
  +\sum_{n=0}^\infty\frac{g_s^n\phi^n}{n!}D^n F(t)\right),
\end{align}
one can make the following interpretation:
$\cF$ is evaluated as the sum
of all possible connected graphs
(allowing loops)
consisting of the trivalent vertex, the $n$-point
vertices $(n\ge 0)$ and the edge,
to which factors $g_s$, $g_s^nD^n F$ and $S$
are assigned respectively.
The symmetry factor for a graph $\Gamma$ is
simply given by $|{\rm Aut}(\Gamma)|^{-1}$,
in the same way as in section~\ref{sec:part}.

We have checked \eqref{eq:masterformula}
by using the explicit data of $\cF_g$ and $F_g$ for small $g$.
As a further consistency check, let us verify that
$\cZ$ given by \eqref{eq:masterformula}
indeed satisfies the holomorphic anomaly equation \eqref{eq:HAE_Z}.
To begin with, let us introduce a formal differential operator
$\tD:=-(\partial_t)_S$,
which is a partial derivative with respect to $t$
holding $S$ constant. We distinguish it from $D=-\partial_t$,
which acts on both $t$ and $S$.
$D$ is expressed in terms of $\tD$ as
\begin{align}
D=S^2\partial_S +\tD.
\end{align}
In other words, we treat $t$ and $S$ as independent variables
and $D_t$ does not act on $S$
\begin{equation}
\begin{aligned}
 D_tf(t)=-\del_t f(t),\quad D_tS=0.
\end{aligned} 
\end{equation}
By plugging \eqref{eq:masterformula} into \eqref{eq:HAE_Z},
the left hand side of \eqref{eq:HAE_Z} is
\begin{equation}
\begin{aligned}
 (2\partial_S-S^{-1})\cZ
&=\int_{-\infty}^\infty d\phi\,
 e^{-\frac{\phi^2}{2S}+\frac{g_s\phi^3}{6}}
 \left(\frac{\phi^2}{S^2}-\frac{1}{S}\right)
 Z(t-g_s\phi)\\
&=\int_{-\infty}^\infty d\phi\,
 e^{-\frac{\phi^2}{2S}+\frac{g_s\phi^3}{6}+g_s\phi\tD}
 \left(\frac{\phi^2}{S^2}-\frac{1}{S}\right)
 Z(t).
\end{aligned} 
\end{equation}
On the other hand, the right hand side of \eqref{eq:HAE_Z} is
\begin{align}
g_s^2(D+S)D\cZ
=\int_{-\infty}^\infty d\phi\,
 e^{-\frac{\phi^2}{2S}+\frac{g_s\phi^3}{6}+g_s\phi\tD}
 g_s^2\left(\tD+S+\frac{\phi^2}{2}\right)
 \left(\tD+\frac{\phi^2}{2}\right)
 Z(t),
\end{align}
where we have used the identity
\begin{align}
e^{\frac{\phi^2}{2S}}S^2\partial_S e^{-\frac{\phi^2}{2S}}
=S^2\partial_S+\frac{\phi^2}{2}.
\end{align}
One can easily show that the difference of the left and right
hand sides of \eqref{eq:HAE_Z} is a total derivative
\begin{align}
\begin{aligned}
&\hspace{-1em}(2\partial_S-S^{-1})\cZ-g_s^2(D+S)D\cZ\\
&=\int_{-\infty}^\infty d\phi
 \frac{d}{d\phi}\left\{
 e^{-\frac{\phi^2}{2S}+\frac{g_s\phi^3}{6}+g_s\phi\tD}
 \left[-\frac{\phi}{S}-g_s\left(\tD+S+\frac{\phi^2}{2}\right)\right]
 Z(t)\right\}.
\end{aligned}
\end{align}
By performing a power series expansion in $g_s$,
one can see that
the boundary contribution vanishes order by order.
Hence, the holomorphic anomaly equation \eqref{eq:HAE_Z} is satisfied.

%%%
\subsection{Large $t$ regime}
%%%

When $t$ is large, one can study the partition function $\cZ$
by means of the power series expansion in $Q=e^{-t}$.
This is done by plugging the fermionic representation
\eqref{eq:Zfermirep} into \eqref{eq:masterformula}:
\begin{align}
\label{eq:cZfermirep}
\begin{aligned}
\cZ
 &=\int_{-\infty}^\infty d\phi\,
 e^{-\frac{\phi^2}{2S}+\frac{g_s\phi^3}{6}}
(Qe^{g_s\phi})^{-\frac{1}{24}}\\
&\hspace{9mm}\times
 \oint\frac{dx}{2\pi\ri x}
 \prod_{p\in\bbZ_{\ge 0}+\hf}
 \left(1+x(Qe^{g_s\phi})^p e^{g_s p^2/2}\right)
 \left(1+x^{-1}(Qe^{g_s\phi})^p e^{-g_s p^2/2}\right).
\end{aligned}
\end{align}
Expanding this expression in $Q$ one can obtain
the small $Q$ expansion of $\cZ$ up to any order.

In particular, in the limit $t\to\infty$, i.e.~$Q=0$,
the partition function is simply given by the Airy integral
\begin{align}
\label{eq:cZlarget}
\lim_{t\to\infty}\h{\cZ}
 &=\frac{1}{\sqrt{2\pi S}}\int_{-\infty}^\infty d\phi\,
 e^{-\frac{g_s\phi}{24}-\frac{\phi^2}{2S}+\frac{g_s\phi^3}{6}},
\end{align}
where $\h{\cZ}$ is defined in \eqref{eq:hatZ}
and we have treated $t$ and $S$ as independent variables.
Note that the Airy function also 
appears in a certain limit of topological string partition function
\cite{Alim:2015qma} and the all-genus resummation of 
the free energy of ABJM theory on $S^3$
\cite{Fuji:2011km,Marino:2011eh}.
We should stress that 
the integral transformation \eqref{eq:masterformula}
defines a mapping between the bosonic and the fermionic partition
functions without assuming a particular limit.
Our formula \eqref{eq:masterformula} is very reminiscent of the
transformation appearing in the Fermi gas formalism 
\cite{Marino:2011eh,Grassi:2014zfa}.
It would be interesting to understand the relation to
\cite{Marino:2011eh,Grassi:2014zfa} better.

%%%
\subsection{Small $t$ regime}
%%%

One can study the small $t$ behavior of the free energy
by using the modular transformation \eqref{eq:modular}.
The Eisenstein series transform as 
\begin{align}
\begin{aligned}
\hE_2(\tau)
 &=\frac{1}{\tau^2}E_2\left(-\frac{1}{\tau}\right)
   +\frac{12}{t}-\frac{12}{t+\bar{t}}\,,\\
E_{2n}(\tau)
 &=\frac{1}{\tau^{2n}}E_{2n}\left(-\frac{1}{\tau}\right)
\qquad (n\ge 2).
\end{aligned}
\end{align}
Therefore, taking $t$ small and setting $\bar{t}=0$,
one has
\begin{align}
\label{eq:Ensmallt}
\begin{aligned}
\hE_2(\tau)
 &=\frac{1}{\tau^2}
   \left[1+{\cal O}\Bigl(e^{-\frac{4\pi^2}{t}}\Bigr)\right],\\
E_{2n}(\tau)
 &=\frac{1}{\tau^{2n}}
   \left[1+{\cal O}\Bigl(e^{-\frac{4\pi^2}{t}}\Bigr)\right]
\qquad (n\ge 2).
\end{aligned}
\end{align}
In this regime one also has
\begin{align}
S=\frac{1}{t},
\end{align}
which follows from \eqref{eq:Sdef} with $\bar{t}=0$.

Taking this into account,
let us regard $\cF_g$ as a polynomial
in the generators $\hE_2$, $E_4$, $E_6$, $S$.
One immediately finds that
$\cF_g$ is of weight $6g-6$, where
weight $2,4,6,2$ are assigned to $\hE_2,E_4,E_6,S$ respectively.
Let us express it as
\begin{align}
\label{eq:Fgpoly}
\cF_g=\sum_{k=0}^{3g-3}P_{g,k}(\hE_2,E_4,E_6)S^k\qquad(g\ge 2).
\end{align}
$P_{g,k}$ is a polynomial of weight $6g-2k-6$. Note that
\begin{align}
P_{g,0}\Big|_{\hE_2=E_2}=F_g.
\end{align}

Based on the explicit data of $\cF_g$ up to high genus
we find the following two conjectural properties.
A peculiar feature of $\cF_g$ is that higher powers of $S$
vanish when written in the basis of $S$ and $\hE_2$
\begin{align}
\label{eq:Pgk0}
P_{g,k}=0\quad\mbox{for}\qquad  k\ge 2g-2.
\end{align}
See \eqref{eq:cF2-hE2},
\eqref{eq:F3resultshE2} and \eqref{eq:F4resultshE2}
for the examples of this property of $\cF_g$ for $g=2,3,4$.
Moreover, if we set
\begin{align}
\label{eq:specialEn}
\hE_2=E_4=E_6=1,
\end{align}
we find
\begin{align}
\label{eq:Psmallt}
P_{g,k}\Big|_{\hE_2=E_4=E_6=1}
=\left\{
\begin{array}{ll}
\displaystyle
\frac{\sqrt{\pi}(1-2^{2g-1})B_{2g}}{2(2g)!\Gamma(5/2-2g)}
&\quad\mbox{for}\quad k= 2g-3,\\[3ex]
0&\quad\mbox{otherwise},
\end{array}
\right.
\end{align}
where $B_k$ is the Bernoulli number (see Appendix~\ref{app:convention}).
In other words,
\begin{align}
\cF_g\Big|_{\hE_2=E_4=E_6=1}
 =\frac{\sqrt{\pi}(1-2^{2g-1})B_{2g}}{2(2g)!\Gamma(5/2-2g)}S^{2g-3}
 \qquad(g\ge 2).
\end{align}

One can now evaluate the behavior of $\cF_g$
in the small $t$ regime as follows.
By plugging \eqref{eq:Ensmallt} into \eqref{eq:Fgpoly},
one sees that all $P_{g,k}$ except for $k=2g-3$ vanish
up to the ${\cal O}(e^{-4\pi^2/t})$ corrections.
The only remaining $P_{g,2g-3}$ is of weight $2g$ and thus it gets
the overall factor $\tau^{-2g}$
along with the value in \eqref{eq:Psmallt}.
Hence we have
\begin{align}
\begin{aligned}
\cF_g\Big|_{t\ll 1,\bar{t}=0}
 &=\tau^{-2g}\cF_g\Big|_{\hE_2=E_4=E_6=1,S=t^{-1}}
   +{\cal O}\Bigl(e^{-\frac{4\pi^2}{t}}\Bigr)\\[1ex]
 &=\frac{\sqrt{\pi}(2^{2g-1}-1)\zeta(2g)}{\Gamma(5/2-2g)}
   t^{3-4g}
   +{\cal O}\Bigl(e^{-\frac{4\pi^2}{t}}\Bigr)
 \qquad(g\ge 2).
\end{aligned}
\label{eq:Fg-small}
\end{align}
Here we have used the relation between $B_{2g}$ and $\zeta(2g)$
in \eqref{eq:zeta-Bk}.
As one can see from \eqref{eq:tdef},
the limit $t\to 0$ corresponds to the weak coupling/small area limit.
$F_g$ in this limit was studied previously
\cite{Rudd:1994ta,Griguolo:2004jp,Griguolo:2004uz}.
Remarkably, here we see that
$\cF_g$ has a much simpler limit than that of $F_g$.

It is known that
the polylogarithm function 
${\rm Li}_s(z):=\sum_{k=1}^\infty z^k/k^s$,
or the Fermi-Dirac integral of order $s-1$,
has the asymptotic expansion of the form
(see e.g.~\cite{Griguolo:2004jp})
\begin{align}
\begin{aligned}
 -{\rm Li}_s(-e^\mu)
&=\frac{1}{\Gamma(s)}\int_0^\infty 
  \frac{\varepsilon^{s-1}d\varepsilon}{e^{\varepsilon-\mu}+1}\\
&=2\sum_{k=0}^\infty
  \frac{(1-2^{1-2k})\zeta(2k)}{\Gamma(s+1-2k)}
  \mu^{s-2k}+{\cal O}(e^{-\mu}).
\end{aligned}
\end{align}
By using this, $\cF$ in the regime
\begin{align}
\label{eq:smalltcond}
\mu\equiv\frac{t^2}{2g_s}\gg 1,\qquad
t\ll 1,\qquad \bar{t}=0
\end{align}
is expressed as
\begin{align}
\label{eq:Fsmallt}
\begin{aligned}
\cF+\frac{t^3}{3g_s^2}
 &=-\sqrt{\frac{\pi}{2g_s}}{\rm Li}_{3/2}(-e^\mu)\\
 &=\sqrt{\frac{2}{g_s}}\int_0^\infty
     \frac{\varepsilon^{1/2}d\varepsilon}{e^{\varepsilon-\mu}+1},
\end{aligned}
\end{align}
where we have ignored the ${\cal O}(e^{-\mu})$
and ${\cal O}(e^{-4\pi^2/t})$ corrections.
This is in accordance with the result of \cite{Griguolo:2004jp},
where the appearance of the Fermi-Dirac integral
in the weak coupling limit was observed
by a quite different approach.
Also, it is interesting to observe that
the non-perturbative correction $e^{-\mu}=e^{-t^2/2g_s}$
agrees with the one found in the resurgence analysis
in \cite{Okuyama:2018clk}. 

Note that the second term on the left hand side of \eqref{eq:Fsmallt}
can be thought of as the genus-zero free energy $\cF_0(t)=t^3/3$
in the bosonic theory, which differs
from that in the fermionic theory $F_0(t)=-t^3/6$
\cite{Vafa:2004qa}.
If we include the genus-zero contribution, the exponential factor of 
\eqref{eq:masterformula}
with $S=1/t$ becomes
\begin{equation}
\begin{aligned}
 \frac{\cF_0(t)}{g_s^2}-\frac{t\phi^2}{2}+\frac{g_s\phi^3}{6}
=\frac{F_0(t-g_s\phi)}{g_s^2}+\frac{\mu(t-g_s\phi)}{g_s}.
\end{aligned}
\end{equation}
Thus, after a change of integration variable $\phi\to -(\phi-t/g_s)$, 
\eqref{eq:masterformula} reduces 
to a very simple expression in the regime of \eqref{eq:smalltcond}
\begin{equation}
\begin{aligned}
\exp\left(-\sqrt{\frac{\pi}{2g_s}}{\rm Li}_{3/2}(-e^\mu)\right)
=\int_{-\infty}^\infty d\phi\, e^{\mu\phi}
 \exp\left(\sum_{g=0}^\infty g_s^{2g-2}F_g(g_s\phi)\right).
\end{aligned} 
\end{equation}

If we neglect the 
non-perturbative corrections of order $\mathcal{O}(e^{-1/t})$,
$\cF_g$ becomes a polynomial in $1/t$.
Such polynomial part of $\cF_g$ can be determined recursively
by solving the holomorphic anomaly equation \eqref{eq:HAE_Fg}
together with \eqref{eq:Fg-small} as the boundary condition at $S=1/t$
to fix the integration constant, also known as the holomorphic
ambiguity. Let us demonstrate this procedure for $\cF_2$ as an
example. We start with the small $t$ behavior of the genus-one free
energy \eqref{eq:cF1}
\begin{equation}
\begin{aligned}
 \cF_1(t,S)=\hf\ln S+\hf\ln t+\frac{\pi^2}{6t}.
\end{aligned} 
\label{eq:small-F1}
\end{equation}
Here and below we ignore the $\mathcal{O}(e^{-1/t})$ corrections.
Plugging \eqref{eq:small-F1}
into the holomorphic anomaly equation \eqref{eq:HAE_Fg} at $g=2$,
we find $\cF_2(t,S)$ up to a holomorphic ambiguity $F_2(t)$
\begin{equation}
\begin{aligned}
 \cF_2(t,S)
 =\frac{5}{24}S^3+ \left(\frac{\pi ^2}{12 t^2}-\frac{1}{4 t}\right)S^2
 + \left(\frac{\pi^4}{72 t^4}+\frac{\pi ^2}{12 t^3}
        -\frac{1}{8 t^2}\right)S
 +F_2(t).
\end{aligned} 
\end{equation}
By imposing the boundary condition at $S=1/t$ \eqref{eq:Fg-small}
\begin{equation}
\begin{aligned}
 \cF_2(t,S=1/t)=\frac{7 \pi ^4}{120 t^5},
\end{aligned} 
\end{equation}
$F_2(t)$ is fixed as
\begin{equation}
\begin{aligned}
 F_2(t)=\frac{2 \pi ^4}{45 t^5}-\frac{\pi ^2}{6
   t^4}+\frac{1}{6 t^3}.
\end{aligned} 
\end{equation}
This agrees with the known small $t$ behavior of $F_2(t)$
\cite{Rudd:1994ta,Griguolo:2004jp,Griguolo:2004uz}.
In a similar manner, one can compute the
higher genus free energy in the small $t$ regime  by
solving the holomorphic anomaly equation recursively.
We should stress that the polynomial-in-$1/t$ part of
the holomorphic ambiguity 
is completely fixed by the boundary condition \eqref{eq:Fg-small}.

%%%
\subsection{Some remarks on determining $\cF_g$}
%%%

As we have seen above, the peculiar features
\eqref{eq:Pgk0} and \eqref{eq:Psmallt}
correspond to the boundary condition of $\cF$ at $t=0$.
In the theory of topological strings,
it is well known that the holomorphic anomaly equation
does not determine the higher genus amplitudes completely,
leaving holomorphic ambiguities.
It is also known that the ambiguities can be
partially (or sometimes completely) removed
by exploiting the polynomial structure
\cite{Yamaguchi:2004bt, Alim:2007qj}
and some boundary conditions at special points
of the moduli space.
While in the present case
we have several ways to determine $\cF_g$ without ambiguity,
it is still interesting to see to what extent
the boundary conditions \eqref{eq:Pgk0} and \eqref{eq:Psmallt}
constrain the form of $\cF_g$.

In what follows let us forget everything
and assume only that
$\cF_g$ has the polynomial structure \eqref{eq:Fgpoly}
and satisfies the holomorphic anomaly equation \eqref{eq:HAE_Fg}.
For given $g\ (\ge 2)$,
the holomorphic anomaly equation completely determines
the form of $P_{g,k}$ for $k>0$, given the forms of
the lower genus amplitudes $\cF_h\ (1\le h\le g-1)$.
Therefore, the only undetermined polynomial is
$P_{g,0}$, which is a quasi modular form of weight $6g-6$.
It consists of $\frac{3g^2}{4}\ \left(\frac{3g^2+1}{4}\right)$
monomials in $\hE_2,E_4,E_6$ when $g$ is even (odd).
We observe that
the boundary conditions \eqref{eq:Pgk0} and \eqref{eq:Psmallt}
partially determine the unknown coefficients.
The number of determined coefficients increases
slightly faster than linearly as $g$ grows.
(For instance,
30 out of 75 coefficients are fixed at $g=10$,
78 out of 300 coefficients are fixed at $g=20$.)

To remove the ambiguities completely,
one way to go further is to evaluate
the higher order corrections to
the expression \eqref{eq:Fsmallt}
and obtain more boundary conditions at $t=0$.
We have the impression that this requires rather intricate analysis.
Alternatively,
instead of/together with \eqref{eq:Pgk0} and \eqref{eq:Psmallt},
one can impose boundary conditions at $t=\infty$.
Such conditions are obtained as many as one wants
by expanding \eqref{eq:cZfermirep} or \eqref{eq:Zfermirep} in $Q$.
As we mentioned in section~\ref{sec:part}, however,
imposing these conditions are essentially equivalent to
solving the recursion relation of \cite{Okuyama:2018clk},
which is much more efficient in determining
$F_g=P_{g,0}\big|_{\hE_2=E_2}$.
We have the impression that solving
the holomorphic anomaly equation \eqref{eq:HAE_Fg}
together with the recursion relation of \cite{Okuyama:2018clk}
is the most efficient way to obtain $\cF_g$
up to given~$g$.

%%%%%%%%%%%%%%%%%%%%%%%%%%%%%%%%%%%%%%%%%%%%%%%%%%%%%%%%%%%%%%%%%%%%%%%%
\section{Conclusions and outlook \label{sec:discussion}}
%%%%%%%%%%%%%%%%%%%%%%%%%%%%%%%%%%%%%%%%%%%%%%%%%%%%%%%%%%%%%%%%%%%%%%%%

In this paper we have clarified how holomorphic anomaly
arises in the partition function of 2d $U(N)$ Yang-Mills theory
on a torus and proposed a natural anti-holomorphic deformation
of the partition function.
Our construction is based on the chiral boson interpretation
of the partition function.
We have pointed out that
consistent recovery of the $\bar{\tau}$-dependence is achieved
by simply ignoring the winding mode contribution.
As a result,
the deformed partition function $\cZ$ defined in this way
contains both $\hE_2$ and $E_2$.
$\hE_2$ always appears through the propagator $G$ in \eqref{eq:prop_z}
and contains the $\bar{\tau}$-dependence
while $E_2$ emerges from period integrals
such as \eqref{eq:pintwp} and should not be replaced with $\hE_2$.
We have demonstrated in detail how to calculate the deformed
free energy $\cF_g$ and $\OPIF_g$.
They are calculated by evaluating
all connected and all 1PI diagrams with $2g-2$ vertices respectively.

We have then identified the precise form of the holomorphic
anomaly equation:
\eqref{eq:HAE_F} for $\cF$,
\eqref{eq:HAE_Z} for $\cZ=\exp\cF$
and \eqref{eq:HAE_1PIF} for $\OPIF$.
We have observed that \eqref{eq:HAE_F} for $\cF$ is very reminiscent
of the traditional BCOV holomorphic anomaly equation.
On the other hand, \eqref{eq:HAE_1PIF} for $\OPIF$
has a rather unconventional form.
However, $\cF$ and $\OPIF$ are related as in \eqref{eq:F-F1PIrel}
and using this relation
we have verified by means of the genus expansion
that two holomorphic anomaly equations
\eqref{eq:HAE_F} and \eqref{eq:HAE_1PIF} are equivalent.

Finally, we have conjectured a closed analytic expression
for the deformed partition function $\cZ$:
It is expressed in terms of the undeformed partition function $Z$
as in \eqref{eq:masterformula}
or as the free fermion representation in \eqref{eq:cZfermirep}.
We have also studied the behavior of $\cZ$
both in the cases of large and small $t$.
In the limit of $t\to\infty$ the partition function
$\cZ$ becomes a mere Airy integral in \eqref{eq:cZlarget}.
On the other hand, in the limit of small $t$ 
with some other conditions \eqref{eq:smalltcond},
the free energy $\cF$ is expressed in terms of
a Fermi-Dirac integral as in \eqref{eq:Fsmallt}.
This small $t$ result of $\cF$
should be compared with the small area limit of $F$.
It turns out that $\cF$ becomes simpler than $F$
due to the occurrence of the
drastic cancellations \eqref{eq:Pgk0} and \eqref{eq:Psmallt}.
We think that this provides us with another nontrivial
support for our anti-holomorphic deformation.

In this paper we have made several conjectures,
which leave room for further investigation.
Our main conjecture is the
closed analytic expression \eqref{eq:masterformula} for $\cZ$.
While we have given it a diagrammatic interpretation,
the expression is still mysterious and we would like to have
a better understanding. For example, it would be very nice if
the form of \eqref{eq:masterformula} is understood directly from
the original path integral \eqref{eq:bosonicZ}.
\eqref{eq:masterformula}
is reminiscent of the relation between the canonical and 
the grand canonical partition functions
in the Fermi gas formalism in \cite{Marino:2011eh,Grassi:2014zfa}.
Also, the appearance of the Airy integral and the Fermi-Dirac integral
suggests a possible connection to the Fermi gas formalism,
which deserves further investigation.

As mentioned above, we have verified that
the two holomorphic anomaly equations
\eqref{eq:HAE_F} for $\cF$ and \eqref{eq:HAE_1PIF} for $\OPIF$
are equivalent. It would be interesting to prove the equivalence.
Another related question is whether
$\OPIF$ also admits a closed analytic expression similar to
\eqref{eq:masterformula} for $\cZ$.
We have discussed the similarity between \eqref{eq:HAE_F}
and the traditional BCOV holomorphic anomaly equation.
It would be interesting to 
understand the more precise relation between them.

In our previous paper \cite{Okuyama:2018clk}, 
we have studied the non-perturbative correction
$\mathcal{O}(e^{-1/g_s})$ 
in the genus expansion of partition function $Z$,
and found evidence that the contribution of baby universes 
advocated in \cite{Dijkgraaf:2005bp}
is not included in 
the partition function of 2d Yang-Mills theory.
To gain more confidence on the absence of
baby universes in the partition function of 2d Yang-Mills theory,
it would be interesting to study the trans-series solution of 
the holomorphic anomaly equation \eqref{eq:HAE_F} along the lines of
\cite{Santamaria:2013rua}.
We leave this as an interesting future problem.

\vskip8mm
\acknowledgments
We would like to thank Ricardo Schiappa for correspondence and
discussion.
This work was supported in part by JSPS KAKENHI Grant Nos. 
26400257  and 16K05316,
and JSPS Japan-Russia Research Cooperative Program.

%%%%%%%%%%%%%%%%%%%%%%%%%%%%%%%%%%%%%%%%%%%%%%%%%%%%%%%%%%%%%%%%%%%%%%%%
\appendix
%%%%%%%%%%%%%%%%%%%%%%%%%%%%%%%%%%%%%%%%%%%%%%%%%%%%%%%%%%%%%%%%%%%%%%%%
\section{\mathversion{bold}Calculation of free energy at $g=3,4$
         \label{app:cF34}}
%%%%%%%%%%%%%%%%%%%%%%%%%%%%%%%%%%%%%%%%%%%%%%%%%%%%%%%%%%%%%%%%%%%%%%%%

In this appendix we calculate $\cF_g$ and $\OPIF_g$ at $g=3,4$.
As shown in Figure~\ref{Fig:Feynman3},
there are five diagrams that contribute to $\cF_3$.
The corresponding integrals are explicitly written as
\begin{align}
\begin{aligned}
\cF_3^{(1)}
 &:=\frac{1}{16}\oint G_{12}^2G_{34}^2G_{13}G_{24},\\
\cF_3^{(2)}
 &:=\frac{1}{24}\oint G_{12}G_{13}G_{14}G_{23}G_{24}G_{34},\\
\cF_3^{(3)}
 &:=\frac{1}{8}\oint G_{12}^2G_{13}G_{23}G_{34}G_{44},\\
\cF_3^{(4)}
 &:=\frac{1}{16}\oint G_{11}G_{12}G_{23}^2G_{34}G_{44},\\
\cF_3^{(5)}
 &:=\frac{1}{48}\oint G_{11}G_{22}G_{33}G_{14}G_{24}G_{34}.
\end{aligned}
\end{align}
Here, we have used the abbreviated notation
\begin{align}
G_{12}\equiv G(z_1,z_2)=G(x_1/x_2)
\end{align}
and suppressed integration variables from the period integral.
\begin{figure}[t]
\begin{equation*}
\begin{array}
{c@{\hspace{3em}}c@{\hspace{3em}}c@{\hspace{2.5em}}c@{\hspace{2.5em}}c}
\begin{tikzpicture}
\draw[line width=1pt](0,0)--(1,0);
\draw[line width=1pt](0,1)--(1,1);
\draw[line width=1pt](0,.5)ellipse(.08 and .5);
\draw[line width=1pt](1,.5)ellipse(.08 and .5);
\filldraw(0,0)circle(.08);
\filldraw(1,0)circle(.08);
\filldraw(0,1)circle(.08);
\filldraw(1,1)circle(.08);
\end{tikzpicture}
&
\begin{tikzpicture}
\draw[line width=1pt](0,0)--(1,0)--(1,1)--(0,1)--(0,0)--(1,1);
\draw[line width=1pt](1,0)--(.55,.45);
\draw[line width=1pt](0,1)--(.45,.55);
\filldraw(0,0)circle(.08);
\filldraw(1,0)circle(.08);
\filldraw(0,1)circle(.08);
\filldraw(1,1)circle(.08);
\end{tikzpicture}
&
\begin{tikzpicture}
\draw[line width=1pt](1,.5)--(1.6,0.5);
\draw[line width=1pt](.5,.5)circle(.5);
\draw[line width=1pt](.5,0)--(.5,1);
\draw[line width=1pt](1.9,.5)circle(.3);
\filldraw(.5,0)circle(.08);
\filldraw(.5,1)circle(.08);
\filldraw(1,.5)circle(.08);
\filldraw(1.6,.5)circle(.08);
\end{tikzpicture}
&
\raisebox{.27cm}{
\begin{tikzpicture}
\draw[line width=1pt](.6,.5)--(1.2,0.5);
\draw[line width=1pt](1.8,.5)--(2.4,0.5);
\draw[line width=1pt](.3,.5)circle(.3);
\draw[line width=1pt](1.5,.5)circle(.3);
\draw[line width=1pt](2.7,.5)circle(.3);
\filldraw(.6,.5)circle(.08);
\filldraw(1.2,.5)circle(.08);
\filldraw(1.8,.5)circle(.08);
\filldraw(2.4,.5)circle(.08);
\end{tikzpicture}
}
&
\begin{tikzpicture}
\draw[line width=1pt](1,1)--(1,1.6);
\draw[line width=1pt](1,1)--(.48,.7);
\draw[line width=1pt](1,1)--(1.52,.7);
\draw[line width=1pt](1,1.9)circle(.3);
\draw[line width=1pt](.22,.55)circle(.3);
\draw[line width=1pt](1.78,.55)circle(.3);
\filldraw(1,1)circle(.08);
\filldraw(1,1.6)circle(.08);
\filldraw(.48,.7)circle(.08);
\filldraw(1.52,.7)circle(.08);
\end{tikzpicture}
\\[.5ex]
\cF_3^{(1)}&\cF_3^{(2)}&\cF_3^{(3)}&\cF_3^{(4)}&\cF_3^{(5)}
\end{array}
\end{equation*}
\caption{Feynman diagrams for $\cF_3$.
\label{Fig:Feynman3}}
\end{figure}

Among these, $\cF_3^{(1)}$ and $\cF_3^{(2)}$
are 1PI, for which the integrals are nontrivial.
To calculate them, we first evaluate the integrals
in the small $Q$ expansion up to sufficiently high order.
We then make the ans\"atze of the form \eqref{eq:IGamma-ansatz}
and match them with the $Q$ expansion results.
In this way we are able to obtain the following exact results
\begin{align}
\begin{aligned}
\cF_3^{(1)}
 &=\frac{1}{16}S^6-\frac{E_2^2-E_4}{1152}S^4
  -\frac{3E_2^4-22E_2^2E_4+32E_2E_6-13E_4^2}{331776}S^2\\
 &\hspace{1em}
  +\frac{E_2^5-4E_2^3E_4+2E_2^2E_6+3E_2E_4^2-2E_4E_6}{497664}S\\
 &\hspace{1em}
  -\frac{3E_2^6-6E_2^4E_4-4E_2^3E_6+3E_2^2E_4^2+12E_2E_4E_6
         -4E_4^3-4E_6^2}
        {35831808},\\
\cF_3^{(2)}
 &=\frac{1}{24}S^6-\frac{E_2^3-3E_2E_4+2E_6}{10368}S^3
  -\frac{E_2^4-6E_2^2E_4+8E_2E_6-3E_4^2}{165888}S^2\\
 &\hspace{1em}
  +\frac{E_2^5-4E_2^3E_4+2E_2^2E_6+3E_2E_4^2-2E_4E_6}{497664}S
  -\frac{E_2^6-3E_2^4E_4+3E_2^2E_4^2-E_4^3}{11943936}.
\end{aligned}
\end{align}
On the other hand, the rest are 1PR and
thus, as shown in \eqref{eq:Ifactorization},
factorize into known pieces
\begin{align}
\begin{aligned}
\cF_3^{(3)}
 &=SD\cF_1D\cF_2^{(1)},\\
\cF_3^{(4)}
 &=\frac{1}{2}S^2\left(D\cF_1\right)^2D^2\cF_1,\\
\cF_3^{(5)}
 &=\frac{1}{6}S^3\left(D\cF_1\right)^3.
\end{aligned}
\end{align}
$D\cF_1$ and $D^2\cF_1$ are given in \eqref{eq:diffcF1}.
$D\cF_2^{(1)}$ is computed from \eqref{eq:F21eval}
by using \eqref{eq:diffformulas} as
\begin{align}
D\cF_2^{(1)}
 &=\frac{1}{4}S^4-\frac{E_2^2-E_4}{576}S^2
   -\frac{E_2^3-3E_2E_4+2E_6}{3456}S
   +\frac{E_2^4-2E_2^2E_4+E_4^2}{41472}.
\end{align}
We thus obtain
\begin{align}
\begin{aligned}
\cF_3
 &=\cF_3^{(1)}+\cF_3^{(2)}+\cF_3^{(3)}+\cF_3^{(4)}+\cF_3^{(5)}\\[1ex]
 &=\frac{5}{16}S^6-\frac{5E_2}{192}S^5-\frac{3E_2^2-5E_4}{2304}S^4
 -\frac{9E_2^3-48E_2E_4+40E_6}{82944}S^3\\
 &\hspace{1em}
 +\frac{2E_2^4+15E_2^2E_4-40E_2E_6+23E_4^2}{331776}S^2\\
 &\hspace{1em}
 +\frac{3E_2^5-14E_2^3E_4+8E_2^2E_6+11E_2E_4^2-8E_4E_6}{995328}S\\
 &\hspace{1em}
 -\frac{6E_2^6-15E_2^4E_4-4E_2^3E_6+12E_2^2E_4^2+12E_2E_4E_6
        -7E_4^3-4E_6^2}
       {35831808},\\[1ex]
\OPIF_3
 &=\cF_3^{(1)}+\cF_3^{(2)}\\[1ex]
 &=\frac{5}{48}S^6-\frac{E_2^2-E_4}{1152}S^4
 -\frac{E_2^3-3E_2E_4+2E_6}{10368}S^3\\
 &\hspace{1em}
 -\frac{5E_2^4-34E_2^2E_4+48E_2E_6-19E_4^2}{331776}S^2\\
 &\hspace{1em}
 +\frac{E_2^5-4E_2^3E_4+2E_2^2E_6+3E_2E_4^2-2E_4E_6}{248832}S\\
 &\hspace{1em}
 -\frac{6E_2^6-15E_2^4E_4-4E_2^3E_6+12E_2^2E_4^2+12E_2E_4E_6
        -7E_4^3-4E_6^2}
       {35831808}.
\end{aligned}
\end{align}
In terms of the basis $\{S,\hE_2,E_4,E_6\}$ they are expressed as
\begin{align}
\label{eq:F3resultshE2}
\begin{aligned}
\cF_3
 &=
-\frac{35\hE_2^3+42\hE_2E_4+16E_6}{27648}S^3
-\frac{58\hE_2^4+33\hE_2^2E_4-40\hE_2E_6-51E_4^2}{331776}S^2\\
 &\hspace{1em}
-\frac{3\hE_2^5-2\hE_2^3E_4-4\hE_2^2E_6-\hE_2E_4^2+4E_4E_6}{331776}S\\
 &\hspace{1em}
-\frac{6\hE_2^6-15\hE_2^4E_4-4\hE_2^3E_6
  +12\hE_2^2E_4^2+12\hE_2E_4E_6-7E_4^3-4E_6^2}{35831808},\\[2ex]
\OPIF_3
 &=
-\frac{17\hE_2^3+27\hE_2E_4+12E_6}{20736}S^3
-\frac{45\hE_2^4+38\hE_2^2E_4-32\hE_2E_6-51E_4^2}{331776}S^2\\
 &\hspace{1em}
-\frac{2\hE_2^5-\hE_2^3E_4-3\hE_2^2E_6-\hE_2E_4^2+3E_4E_6}{248832}S\\
 &\hspace{1em}
-\frac{6\hE_2^6-15\hE_2^4E_4-4\hE_2^3E_6
  +12\hE_2^2E_4^2+12\hE_2E_4E_6-7E_4^3-4E_6^2}{35831808}.
\end{aligned}
\end{align}

\newpage

\begin{figure}[t]
\begin{equation*}
\begin{array}
{c@{\hspace{2.5em}}c@{\hspace{2.5em}}
 c@{\hspace{2.5em}}c@{\hspace{2.5em}}c}
\begin{tikzpicture}
\draw[line width=1pt]
 (.5,0)--(1.5,0)--(2,.866)--(1.5,1.732)--(.5,1.732)--(0,.866)--(.5,0);
\draw[line width=1pt](0,.866)--(2,.866);
\draw[line width=1pt](.5,0)--(.9615,.799);
\draw[line width=1pt](1.0385,.933)--(1.5,1.732);
\draw[line width=1pt](1.5,0)--(1.0385,.799);
\draw[line width=1pt](.9615,.933)--(.5,1.732);
\filldraw(.5,0)circle(.08);
\filldraw(1.5,0)circle(.08);
\filldraw(2,.866)circle(.08);
\filldraw(1.5,1.732)circle(.08);
\filldraw(.5,1.732)circle(.08);
\filldraw(0,.866)circle(.08);
\end{tikzpicture}
&
\begin{tikzpicture}
\draw[line width=1pt]
 (.5,0)--(1.5,0)--(2,.866)--(1.5,1.732)--(.5,1.732)--(0,.866)--(.5,0);
\draw[line width=1pt](0,.866)--(2,.866);
\draw[line width=1pt](.5,0)--(.5,.799);
\draw[line width=1pt](.5,.933)--(.5,1.732);
\draw[line width=1pt](1.5,0)--(1.5,.799);
\draw[line width=1pt](1.5,.933)--(1.5,1.732);
\filldraw(.5,0)circle(.08);
\filldraw(1.5,0)circle(.08);
\filldraw(2,.866)circle(.08);
\filldraw(1.5,1.732)circle(.08);
\filldraw(.5,1.732)circle(.08);
\filldraw(0,.866)circle(.08);
\end{tikzpicture}
&
\begin{tikzpicture}
\draw[line width=1pt](.5,0)--(1.5,0)--(2,.866)--(1.5,1.732);
\draw[line width=1pt](.5,1.732)--(0,.866)--(.5,0);
\draw[line width=1pt](.5,0)--(2,.866);
\draw[line width=1pt](1.5,0)--(1.06,.254);
\draw[line width=1pt](.94,.3233)--(0,.866);
\draw[line width=1pt](1,1.732)ellipse(.5 and .08);
\filldraw(.5,0)circle(.08);
\filldraw(1.5,0)circle(.08);
\filldraw(2,.866)circle(.08);
\filldraw(1.5,1.732)circle(.08);
\filldraw(.5,1.732)circle(.08);
\filldraw(0,.866)circle(.08);
\end{tikzpicture}
&
\raisebox{-.011cm}{
\begin{tikzpicture}
\draw[line width=1pt](1.5,0)--(2,.866)--(1.5,1.732);
\draw[line width=1pt](.5,1.732)--(0,.866)--(.5,0);
\draw[line width=1pt](0,.866)--(2,.866);
\draw[line width=1pt](1,0)ellipse(.5 and .08);
\draw[line width=1pt](1,1.732)ellipse(.5 and .08);
\filldraw(.5,0)circle(.08);
\filldraw(1.5,0)circle(.08);
\filldraw(2,.866)circle(.08);
\filldraw(1.5,1.732)circle(.08);
\filldraw(.5,1.732)circle(.08);
\filldraw(0,.866)circle(.08);
\end{tikzpicture}
}
&
\begin{tikzpicture}
\draw[line width=1pt](.5,0)--(1.5,0);
\draw[line width=1pt](2,.866)--(1.5,1.732);
\draw[line width=1pt](.5,1.732)--(0,.866);
\draw[line width=1pt](1,1.732)ellipse(.5 and .08);
\draw[line width=1pt](.25,.433)
  circle[x radius=.5,y radius=.08,rotate=120];
\draw[line width=1pt](1.75,.433)
  circle[x radius=.5,y radius=.08,rotate=60];
\filldraw(.5,0)circle(.08);
\filldraw(1.5,0)circle(.08);
\filldraw(2,.866)circle(.08);
\filldraw(1.5,1.732)circle(.08);
\filldraw(.5,1.732)circle(.08);
\filldraw(0,.866)circle(.08);
\end{tikzpicture}
\\[.5ex]
72&12&8&16&48
\end{array}
\end{equation*}

\vspace{6ex}
\begin{equation*}
\begin{array}
{c@{\hspace{3em}}c@{\hspace{3em}}c@{\hspace{3em}}c}
\begin{tikzpicture}
\draw[line width=1pt](0,0)--(1,0);
\draw[line width=1pt](0,1)--(1,1);
\draw[line width=1pt](0,.5)ellipse(.08 and .5);
\draw[line width=1pt](1,.5)ellipse(.08 and .5);
\draw[line width=1pt](1.08,0.5)--(1.6,.5);
\draw[line width=1pt](1.9,.5)circle(.3);
\filldraw(0,0)circle(.08);
\filldraw(1,0)circle(.08);
\filldraw(0,1)circle(.08);
\filldraw(1,1)circle(.08);
\filldraw(1.08,.5)circle(.08);
\filldraw(1.6,.5)circle(.08);
\end{tikzpicture}
&
\raisebox{-.011cm}{
\begin{tikzpicture}
\draw[line width=1pt](0,0)--(0,1);
\draw[line width=1pt](1,0)--(1,1);
\draw[line width=1pt](.5,0)ellipse(.5 and .08);
\draw[line width=1pt](.5,1)ellipse(.5 and .08);
\draw[line width=1pt](1,0.5)--(1.6,.5);
\draw[line width=1pt](1.9,.5)circle(.3);
\filldraw(0,0)circle(.08);
\filldraw(1,0)circle(.08);
\filldraw(0,1)circle(.08);
\filldraw(1,1)circle(.08);
\filldraw(1,.5)circle(.08);
\filldraw(1.6,.5)circle(.08);
\end{tikzpicture}
}
&
\begin{tikzpicture}
\draw[line width=1pt](0,0)--(1,0)--(1,1)--(0,1)--(0,0)--(1,1);
\draw[line width=1pt](1,0)--(.55,.45);
\draw[line width=1pt](0,1)--(.45,.55);
\draw[line width=1pt](1,0.5)--(1.6,.5);
\draw[line width=1pt](1.9,.5)circle(.3);
\filldraw(0,0)circle(.08);
\filldraw(1,0)circle(.08);
\filldraw(0,1)circle(.08);
\filldraw(1,1)circle(.08);
\filldraw(1,.5)circle(.08);
\filldraw(1.6,.5)circle(.08);
\end{tikzpicture}
&
\begin{tikzpicture}
\draw[line width=1pt](1,.5)--(1.6,.5);
\draw[line width=1pt](.5,.5)circle(.5);
\draw[line width=1pt](.5,0)--(.5,1);
\draw[line width=1pt](2.1,.5)circle(.5);
\draw[line width=1pt](2.1,0)--(2.1,1);
\filldraw(.5,0)circle(.08);
\filldraw(.5,1)circle(.08);
\filldraw(1,.5)circle(.08);
\filldraw(1.6,.5)circle(.08);
\filldraw(2.1,0)circle(.08);
\filldraw(2.1,1)circle(.08);
\end{tikzpicture}
\\[.5ex]
8&16&8&32
\end{array}
\end{equation*}

\vspace{6ex}
\begin{equation*}
\begin{array}
{c@{\hspace{3em}}c@{\hspace{3em}}c}
\begin{tikzpicture}
\draw[line width=1pt](1,.5)--(1.6,.5);
\draw[line width=1pt](.5,.5)circle(.5);
\draw[line width=1pt](.5,0)--(.5,1);
\draw[line width=1pt](1.9,.5)circle(.3);
\draw[line width=1pt](2.2,.5)--(2.8,.5);
\draw[line width=1pt](3.1,.5)circle(.3);
\filldraw(.5,0)circle(.08);
\filldraw(.5,1)circle(.08);
\filldraw(1,.5)circle(.08);
\filldraw(1.6,.5)circle(.08);
\filldraw(2.2,.5)circle(.08);
\filldraw(2.8,.5)circle(.08);
\end{tikzpicture}
&
\begin{tikzpicture}
\draw[line width=1pt](.3,.5)circle(.3);
\draw[line width=1pt](1.7,.5)circle(.5);
\draw[line width=1pt](3.1,.5)circle(.3);
\draw[line width=1pt](.6,.5)--(1.2,.5);
\draw[line width=1pt](2.2,.5)--(2.8,.5);
\draw[line width=1pt](1.7,0)--(1.7,1);
\filldraw(.6,.5)circle(.08);
\filldraw(1.2,.5)circle(.08);
\filldraw(1.7,0)circle(.08);
\filldraw(1.7,1)circle(.08);
\filldraw(2.2,.5)circle(.08);
\filldraw(2.8,.5)circle(.08);
\end{tikzpicture}
&
\begin{tikzpicture}
\draw[line width=1pt](.3,.1464)circle(.3);
\draw[line width=1pt](1.7,.5)circle(.5);
\draw[line width=1pt](3.1,.1464)circle(.3);
\draw[line width=1pt](.6,.1464)--(1.346,.1464);
\draw[line width=1pt](2.054,.1464)--(2.8,.1464);
\draw[line width=1pt](1.346,.8536) to [out=315,in=225] (2.054,.8536);
\filldraw(.6,.1464)circle(.08);
\filldraw(1.346,.8536)circle(.08);
\filldraw(2.054,.8536)circle(.08);
\filldraw(1.346,.1464)circle(.08);
\filldraw(2.054,.1464)circle(.08);
\filldraw(2.8,.1464)circle(.08);
\end{tikzpicture}
\\[.5ex]
16&16&16
\end{array}
\end{equation*}

\vspace{4ex}
\begin{equation*}
\begin{array}
{c@{\hspace{3em}}c@{\hspace{3em}}c}
\raisebox{.26cm}{
\begin{tikzpicture}
\draw[line width=1pt](1,0)--(1.6,0);
\draw[line width=1pt](.5,0)circle(.5);
\draw[line width=1pt](.5,-.5)--(.5,.5);
\draw[line width=1pt](1.6,0)--(1.9,.5196);
\draw[line width=1pt](1.6,0)--(1.9,-.5196);
\draw[line width=1pt](2.05,.7794)circle(.3);
\draw[line width=1pt](2.05,-.7794)circle(.3);
\filldraw(.5,-.5)circle(.08);
\filldraw(.5,.5)circle(.08);
\filldraw(1,0)circle(.08);
\filldraw(1.6,0)circle(.08);
\filldraw(1.9,.5196)circle(.08);
\filldraw(1.9,-.5196)circle(.08);
\end{tikzpicture}
}
&
\raisebox{1.038cm}{
\begin{tikzpicture}
\draw[line width=1pt](.6,.5)--(1.2,.5);
\draw[line width=1pt](1.8,.5)--(2.4,0.5);
\draw[line width=1pt](3.0,.5)--(3.6,0.5);
\draw[line width=1pt](.3,.5)circle(.3);
\draw[line width=1pt](1.5,.5)circle(.3);
\draw[line width=1pt](2.7,.5)circle(.3);
\draw[line width=1pt](3.9,.5)circle(.3);
\filldraw(.6,.5)circle(.08);
\filldraw(1.2,.5)circle(.08);
\filldraw(1.8,.5)circle(.08);
\filldraw(2.4,.5)circle(.08);
\filldraw(3.0,.5)circle(.08);
\filldraw(3.6,.5)circle(.08);
\end{tikzpicture}
}
&
\begin{tikzpicture}
\draw[line width=1pt](.6,0)--(1.2,0);
\draw[line width=1pt](1.65,.26)--(1.95,.7794);
\draw[line width=1pt](1.65,-.26)--(1.95,-.7794);
\draw[line width=1pt](.3,0)circle(.3);
\draw[line width=1pt](1.5,0)circle(.3);
\draw[line width=1pt](2.1,1.039)circle(.3);
\draw[line width=1pt](2.1,-1.039)circle(.3);
\filldraw(.6,0)circle(.08);
\filldraw(1.2,0)circle(.08);
\filldraw(1.65,.26)circle(.08);
\filldraw(1.65,-.26)circle(.08);
\filldraw(1.95,.7794)circle(.08);
\filldraw(1.95,-.7794)circle(.08);
\end{tikzpicture}
\\[-.5ex]
32&32&48
\end{array}
\end{equation*}

\vspace{4ex}
\begin{equation*}
\begin{array}
{c@{\hspace{4em}}c}
\begin{tikzpicture}
\draw[line width=1pt](.6,0)--(1.2,0);
\draw[line width=1pt](1.8,0)--(2.4,0);
\draw[line width=1pt](.3,0)circle(.3);
\draw[line width=1pt](1.5,0)circle(.3);
\filldraw(.6,0)circle(.08);
\filldraw(1.2,0)circle(.08);
\filldraw(1.8,0)circle(.08);
\filldraw(2.4,0)circle(.08);
\draw[line width=1pt](2.4,0)--(2.7,.5196);
\draw[line width=1pt](2.4,0)--(2.7,-.5196);
\draw[line width=1pt](2.85,.7794)circle(.3);
\draw[line width=1pt](2.85,-.7794)circle(.3);
\filldraw(2.4,0)circle(.08);
\filldraw(2.7,.5196)circle(.08);
\filldraw(2.7,-.5196)circle(.08);
\end{tikzpicture}
&
\begin{tikzpicture}
\draw[line width=1pt](-.3,0)--(-.6,.5196);
\draw[line width=1pt](-.3,0)--(-.6,-.5196);
\draw[line width=1pt](.3,0)--(.6,.5196);
\draw[line width=1pt](.3,0)--(.6,-.5196);
\draw[line width=1pt](-.3,0)--(.3,0);
\draw[line width=1pt](-.75,.7794)circle(.3);
\draw[line width=1pt](-.75,-.7794)circle(.3);
\draw[line width=1pt](.75,.7794)circle(.3);
\draw[line width=1pt](.75,-.7794)circle(.3);
\filldraw(-.6,.5196)circle(.08);
\filldraw(-.6,-.5196)circle(.08);
\filldraw(-.3,0)circle(.08);
\filldraw(.3,0)circle(.08);
\filldraw(.6,.5196)circle(.08);
\filldraw(.6,-.5196)circle(.08);
\end{tikzpicture}
\\[.5ex]
32&128
\end{array}
\end{equation*}
\caption{Feynman diagrams for $\cF_4$.
There are 17 diagrams:
the five hexagonal diagrams on the top line
are 1PI and the others are 1PR.
The number under each diagram shows
the order of the automorphism group of the graph,
which gives the inverse of the symmetry factor.
\label{Fig:Feynman4}}
\end{figure}
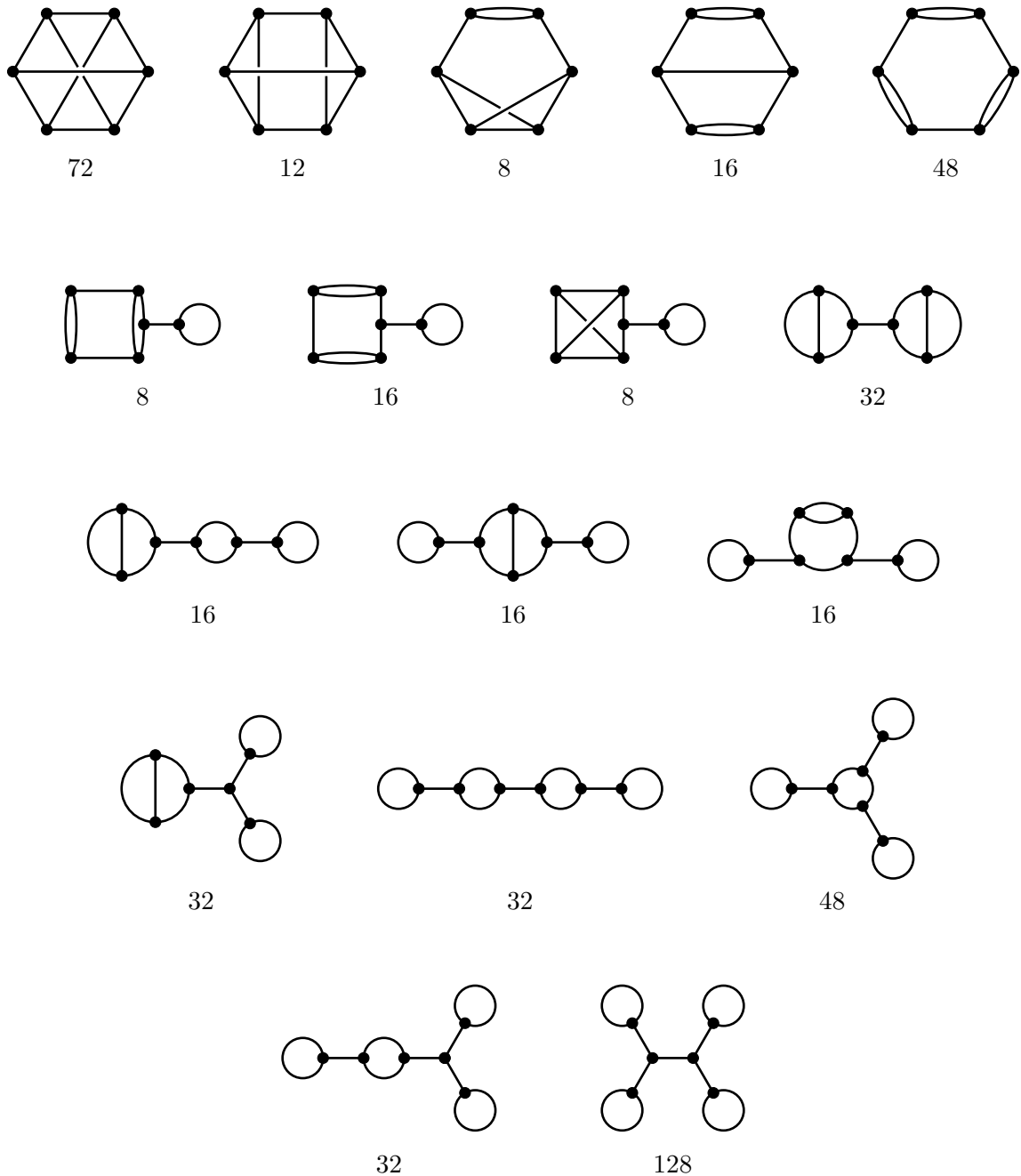

One can calculate $\cF_4$ and $\OPIF_4$ in the same way.
As shown in Figure~\ref{Fig:Feynman4}, there are
17 diagrams that contribute to $\cF_4$.
The first five of them are 1PI while the others are 1PR.
The calculations are tedious but straightforward.
The final results are as follows:
\begin{align}
\begin{aligned}
\cF_4
 &=
\frac{1105}{1152}S^9
-\frac{5}{64}E_2S^8
-\frac{5(7E_2^2-12E_4)}{9216}S^7
-\frac{53E_2^3-285E_2E_4+240E_6}{165888}S^6\\
 &\hspace{1em}
-\frac{31E_2^4-712E_2^2E_4+1360E_2E_6-680E_4^2}{2654208}S^5\\
 &\hspace{1em}
+\frac{15E_2^5+179E_2^3E_4-760E_2^2E_6+910E_2E_4^2-344E_4E_6}{7962624}
  S^4\\
 &\hspace{1em}
+\frac{1}{286654464}(255E_2^6-1290E_2^4E_4-812E_2^3E_6\\
 &\hspace{3em}
  +6393E_2^2E_4^2-6720E_2E_4E_6+1070E_4^3+1104E_6^2)S^3\\
 &\hspace{1em}
+\frac{1}{286654464}(25E_2^7-385E_2^5E_4+696E_2^4E_6+211E_2^3E_4^2\\
 &\hspace{3em}
  -1484E_2^2E_4E_6+725E_2E_4^3+576E_2E_6^2-364E_4^2E_6)S^2\\
 &\hspace{1em}
-\frac{1}{3439853568}
  (71E_2^8-496E_2^6E_4+280E_2^5E_6+886E_2^4E_4^2-400E_2^3E_4E_6\\
 &\hspace{3em}
  -1016E_2^2E_4^3-448E_2^2E_6^2+1592E_2E_4^2E_6-181E_4^4-288E_4E_6^2)S\\
 &\hspace{1em}
+\frac{1}{464380231680}
  (355E_2^9-1395E_2^7E_4-600E_2^6E_6+1737E_2^5E_4^2\\
 &\hspace{3em}
  +4410E_2^4E_4E_6-2145E_2^3E_4^3-1860E_2^3E_6^2-6300E_2^2E_4^2E_6\\
 &\hspace{3em}
  +3600E_2E_4^4+4860E_2E_4E_6^2-2238E_4^3E_6-424E_6^3),\\[2ex]
\OPIF_4
 &=
\frac{11}{36}S^9
-\frac{5(E_2^2-E_4)}{2304}S^7
-\frac{5(E_2^3-3E_2E_4+2E_6)}{20736}S^6\\
 &\hspace{1em}
-\frac{5E_2^4-34E_2^2E_4+48E_2E_6-19E_4^2}{165888}S^5\\
 &\hspace{1em}
-\frac{E_2^5-28E_2^3E_4+74E_2^2E_6-69E_2E_4^2+22E_4E_6}{995328}S^4\\
 &\hspace{1em}
+\frac{1}{143327232}(41E_2^6+225E_2^4E_4-1728E_2^3E_6\\
 &\hspace{3em}
  +3219E_2^2E_4^2-2400E_2E_4E_6+307E_4^3+336E_6^2)S^3\\
 &\hspace{1em}
+\frac{1}{35831808}(7E_2^7-71E_2^5E_4+104E_2^4E_6+49E_2^3E_4^2\\
 &\hspace{3em}
  -208E_2^2E_4E_6+87E_2E_4^3+72E_2E_6^2-40E_4^2E_6)S^2\\
 &\hspace{1em}
-\frac{1}{286654464}
  (7E_2^8-46E_2^6E_4+24E_2^5E_6+80E_2^4E_4^2-32E_2^3E_4E_6\\
 &\hspace{3em}
  -90E_2^2E_4^3-40E_2^2E_6^2+136E_2E_4^2E_6-15E_4^4-24E_4E_6^2)S\\
 &\hspace{1em}
+\frac{1}{464380231680}
  (355E_2^9-1395E_2^7E_4-600E_2^6E_6+1737E_2^5E_4^2\\
 &\hspace{3em}
  +4410E_2^4E_4E_6-2145E_2^3E_4^3-1860E_2^3E_6^2-6300E_2^2E_4^2E_6\\
 &\hspace{3em}
  +3600E_2E_4^4+4860E_2E_4E_6^2-2238E_4^3E_6-424E_6^3).
\end{aligned}
\end{align}
In terms of the basis $\{S,\hE_2,E_4,E_6\}$ they are expressed as
\begin{align}
\label{eq:F4resultshE2}
\begin{aligned}
\cF_4
 &=
\frac{11(175\hE_2^4+420\hE_2^2E_4+320\hE_2E_6+228E_4^2)}{1474560}S^5\\
 &\hspace{1em}
+\frac{875\hE_2^5+1793\hE_2^3E_4
  +244\hE_2^2E_6-1128\hE_2E_4^2-1784E_4E_6}{2654208}S^4\\
 &\hspace{1em}
+\frac{1}{286654464}(10307\hE_2^6+12810\hE_2^4E_4-13804\hE_2^3E_6\\
 &\hspace{3em}
  -31275\hE_2^2E_4^2-9120\hE_2E_4E_6+19674E_4^3+11408E_6^2)S^3\\
 &\hspace{1em}
+\frac{1}{286654464}
  (593\hE_2^7-13\hE_2^5E_4-1504\hE_2^4E_6-1789\hE_2^3E_4^2\\
 &\hspace{3em}
  +2068\hE_2^2E_4E_6+2185\hE_2E_4^3+976\hE_2E_6^2-2516E_4^2E_6)S^2\\
 &\hspace{1em}
+\frac{1}{1146617856}(71\hE_2^8-124\hE_2^6E_4
  -200\hE_2^5E_6-38\hE_2^4E_4^2+656\hE_2^3E_4E_6\\
 &\hspace{3em}
  +148\hE_2^2E_4^3-16\hE_2^2E_6^2
  -904\hE_2E_4^2E_6+167E_4^4+240E_4E_6^2)S\\
 &\hspace{1em}
+\frac{1}{464380231680}
  (355\hE_2^9-1395\hE_2^7E_4-600\hE_2^6E_6+1737\hE_2^5E_4^2\\
 &\hspace{3em}
  +4410\hE_2^4E_4E_6-2145\hE_2^3E_4^3
  -1860\hE_2^3E_6^2-6300\hE_2^2E_4^2E_6\\
 &\hspace{3em}
  +3600\hE_2E_4^4+4860\hE_2E_4E_6^2-2238E_4^3E_6-424E_6^3),\\[2ex]
\OPIF_4
 &=
\frac{1435\hE_2^4+4230\hE_2^2E_4+3600\hE_2E_6+2799E_4^2}{1658880}S^5\\
 &\hspace{1em}
+\frac{487\hE_2^5+1226\hE_2^3E_4
  +300\hE_2^2E_6-681\hE_2E_4^2-1332E_4E_6}{1990656}S^4\\
 &\hspace{1em}
+\frac{1}{143327232}(4185\hE_2^6+6825\hE_2^4E_4-5440\hE_2^3E_6\\
 &\hspace{3em}
  -15021\hE_2^2E_4^2-6048\hE_2E_4E_6+9819E_4^3+5680E_6^2)S^3\\
 &\hspace{1em}
+\frac{1}{71663616}
  (130\hE_2^7+35\hE_2^5E_4-352\hE_2^4E_6-476\hE_2^3E_4^2\\
 &\hspace{3em}
  +460\hE_2^2E_4E_6+571\hE_2E_4^3+260\hE_2E_6^2-628E_4^2E_6)S^2\\
 &\hspace{1em}
+\frac{1}{859963392}
  (50\hE_2^8-79\hE_2^6E_4-152\hE_2^5E_6
  -47\hE_2^4E_4^2+488\hE_2^3E_4E_6\\
 &\hspace{3em}
  +127\hE_2^2E_4^3-4\hE_2^2E_6^2
  -688\hE_2E_4^2E_6+125E_4^4+180E_4E_6^2)S\\
 &\hspace{1em}
+\frac{1}{464380231680}
  (355\hE_2^9-1395\hE_2^7E_4-600\hE_2^6E_6+1737\hE_2^5E_4^2\\
 &\hspace{3em}
  +4410\hE_2^4E_4E_6-2145\hE_2^3E_4^3
  -1860\hE_2^3E_6^2-6300\hE_2^2E_4^2E_6\\
 &\hspace{3em}
  +3600\hE_2E_4^4+4860\hE_2E_4E_6^2-2238E_4^3E_6-424E_6^3).
\end{aligned}
\end{align}

\newpage
~
\newpage
%%%%%%%%%%%%%%%%%%%%%%%%%%%%%%%%%%%%%%%%%%%%%%%%%%%%%%%%%%%%%%%%%%%%%%%%
\section{Convention of special functions and some useful relations
         \label{app:convention}}
%%%%%%%%%%%%%%%%%%%%%%%%%%%%%%%%%%%%%%%%%%%%%%%%%%%%%%%%%%%%%%%%%%%%%%%%

The Weierstrass $\wp$-function
is defined as
\begin{align}
\wp(z;2\omega_1,2\omega_3)
  &:=\frac{1}{z^2}
  +\sum_{(m,n)\in\bbZ^2_{\ne (0,0)}}
  \left[\frac{1}{(z-\Omega_{m,n})^2}
    -\frac{1}{{\Omega_{m,n}}^2}\right],
\end{align}
where $\Omega_{m,n}=2m\omega_1 + 2n\omega_3$.
In this paper we always set
$2\omega_1=2\pi,\ 2\omega_3=2\pi\tau$
and use the following abbreviated notation
\begin{align}
\wp(z):=\wp(z;2\pi,2\pi\tau).
\end{align}
The Dedekind eta function is defined as
\begin{align}
\eta(\tau):=Q^{1/24}\prod_{n=1}^\infty(1-Q^n).
\end{align}
The Eisenstein series are given by
\begin{align}
E_{2n}(\tau)
 =1-\frac{4n}{B_{2n}}\sum_{k=1}^{\infty}\frac{k^{2n-1}Q^k}{1-Q^k}
\end{align}
for $n\in\bbZ_{>0}$. The Bernoulli numbers $B_k$ are defined by
\begin{align}
\frac{x}{e^x-1}=\sum_{k=0}^\infty\frac{B_k}{k!}x^k.
\end{align}
The value of zeta-function at a non-negative even integer is 
given by 
\begin{equation}
\begin{aligned}
\zeta(2k)=\frac{(-1)^{k+1}(2\pi)^{2k}B_{2k}}{2(2k)!}
\qquad (k\in\mathbb{Z}_{\ge 0}).
\end{aligned} 
\label{eq:zeta-Bk}
\end{equation}
We often abbreviate $\eta(\tau),\,E_{2n}(\tau)$ as
$\eta,\,E_{2n}$ respectively.

In the main text we use the following
differentiation formulas
\begin{align}
\label{eq:diffformulas}
\begin{aligned}
DE_2&=\frac{E_2^2-E_4}{12},\quad&
DE_4&=\frac{E_2E_4-E_6}{3},\quad&
DE_6&=\frac{E_2E_6-E_4^2}{2},\\
D\ln\eta&=\frac{E_2}{24},&
DS&=S^2,&&&
\end{aligned}
\end{align}
where $D:=Q\partial_Q=(2\pi\ri)^{-1}\partial_\tau$
and $S:=(4\pi{\rm Im}\tau)^{-1}=(E_2-\hE_2)/12$.

Under the modular $S$-transformation we have
\begin{equation}
\begin{aligned}
E_{2n}(-1/\tau)&=\tau^{2n}E_{2n}(\tau),\quad(n\geq2),\\
E_2(-1/\tau)&=\tau^2 E_2(\tau)+\frac{6\tau}{\pi\ri},\\
S(-1/\tau,-1/\bar{\tau})
 &=\tau^2 S(\tau,\bar{\tau})+\frac{\tau}{2\pi\ri},\\
\hE_2(-1/\tau,-1/\bar{\tau})&=\tau^2 \hE_2(\tau,\bar{\tau}),\\
\eta(-1/\tau)&=\rt{-\ri\tau}\eta(\tau).
\end{aligned} 
\label{eq:modular}
\end{equation}
Here we have regarded $S$ as a function of $\tau$ and $\bar{\tau}$:
 $S=S(\tau,\bar{\tau})=(4\pi\text{Im}\tau)^{-1}$.

%%%%%%%%%%%%%%%%%%%%%%%%%%%%%%%%%%%%%%%%%%%%%%%%%%%%%%%%%%%%%%%%%%%%%%%%

\end{document}